\documentclass[11pt]{article}

\usepackage{slashed}
\usepackage{xcolor}

\def\lsim{\mathrel{\raise.3ex\hbox{$<$\kern-.75em\lower1ex\hbox{$\sim$}}}}
\def\gsim{\mathrel{\raise.3ex\hbox{$>$\kern-.75em\lower1ex\hbox{$\sim$}}}}

\definecolor{red}{rgb}{1.0, 0, 0}

\newcommand{\hc}{\text{h.c.}}

\usepackage{jheppub} 
\usepackage{bm,float}
\usepackage[caption = false]{subfig}
\usepackage[section]{placeins} 
\usepackage{multirow, booktabs, amsthm, xfrac, fix-cm}
\usepackage{url, amsmath, amssymb, color, setspace, tabu}

\def \LC{\epsilon}

\def \h{\mathfrak{H}}

\usepackage{etoolbox}
\makeatletter
\patchcmd{\maketitle}{\@fpheader}{ }{ }{ }
\makeatother

\begin{document}

\title{Low-derivative operators of the Standard Model effective field theory via
Hilbert series methods}

\author{Landon Lehman}

\author{and Adam Martin}
\affiliation{Department of Physics, University of Notre Dame, Notre Dame, IN
46556}
\emailAdd{llehman@nd.edu}
\emailAdd{amarti41@nd.edu}

\date{\today}
\preprint{ }

\abstract{
In this work, we explore an extension of Hilbert series techniques to count
operators that include derivatives. For sufficiently low-derivative operators,
we find an algorithm that gives the number of invariant operators, properly
accounting for redundancies due to the equations of motion and integration by
parts. Specifically, the technique can be applied whenever there is only one
Lorentz invariant for a given partitioning of derivatives among the fields. At
higher numbers of derivatives, equation of motion redundancies can be removed,
but the increased number of Lorentz contractions spoils the subtraction of
integration by parts redundancies. While restricted, this technique is
sufficient to automatically generate the complete set of invariant operators of
the Standard Model effective field theory for dimensions 6 and 7 (for arbitrary
numbers of flavors). At dimension 8, the algorithm does not automatically
generate the complete operator set; however, it suffices for all but five
classes of operators. For these remaining classes, there is a well defined
procedure to manually determine the number of invariants. Using these methods,
we thereby derive the set of 535 dimension-8 $N_f =1$ operators.}
	
\maketitle

In a previous paper~\cite{Lehman:2015via}, we reviewed the Hilbert series
technique and highlighted its application for determining the complete set of
operators in an effective field theory (EFT). We gave several examples, both
using only Standard Model (SM) field content and also using beyond the Standard
Model setups. All of the examples in \cite{Lehman:2015via}, however, were
restricted to non-derivative operators.

Operators with derivatives face two additional complications compared to
operators without derivatives: redundancy due to the equations of motion (EOM),
and redundancy due to integration by parts (IBP). Work on both of these issues
has recently been presented in the context of scalars in $(0+1)$ dimensional
spacetime~\cite{berkeley}. Inspired by this success in lower dimensions, the
goal of this present work is to extend the Hilbert series technique of invariant
counting to operators with derivatives in four spacetime dimensions. 

While a complete method for incorporating derivatives and their associated
issues into Hilbert series for four-dimensional EFTs remains out of grasp, we have
found an algorithm that suffices for operators with low numbers of derivatives.
Exactly what counts as a `low-derivative' operator depends on the fields
involved, but typically it means fewer than two derivatives.  Before introducing
our algorithm to incorporate derivatives into Hilbert series, we first briefly
review the Hilbert series method and more clearly define the derivative issues
in that context.

\section{Introduction and Hilbert series review}
\label{sec:intro}

The Hilbert series is defined as a power series \begin{equation} \h = \sum c_n\,
	t^n, \end{equation} where $t^n$ indicates the invariants involving $n$
objects $t$, and $c_n$ is the number of independent invariants at that order.
The Hilbert series technique has previously been used in wide-ranging
theoretical contexts \cite{Pouliot:1998yv, Benvenuti:2006qr, Dolan:2007rq,
	Gray:2008yu, Hanany:2008sb, Chen:2011wn, Butti:2007jv, Feng:2007ur,
	Forcella:2007wk, Benvenuti:2010pq, Hanany:2012dm,
Rodriguez-Gomez:2013dpa, Dey:2013fea, Hanany:2014hia, Begin:1998hn,
Hanany:2014dia} and in more phenomenologically oriented applications
\cite{Jenkins:2009dy, Hanany:2010vu, Merle:2011vy}, but the issues associated
with derivatives have not yet been addressed. The phenomenological utility of
this abstract mathematical object $\h$ becomes clear once we identify $t$ with
operators of some quantum field theory (QFT) and $n$ as the overall mass
dimension of the operator; $\h$ then gives the set of gauge, global, and Lorentz
invariant operators of mass dimension $n$. Here $c_n$ tells us the number of
independent invariants, but nothing about the values of the coefficients those
invariants may have in the Lagrangian. These details depend on UV physics.

While Hilbert series techniques can be applied to any QFT,\footnote{Provided the
fields transform linearly under the symmetries of the theory, and the symmetry
groups are compact Lie groups.} the application
we are most interested in here is to count invariants of the Standard Model
effective field theory (SMEFT). Given the absence of new light states in the
first run of the LHC, the SMEFT is both a powerful and an agnostic way to bound
new physics. Since the Higgs discovery in 2012 there has been intense work on
the SMEFT on both the theoretical and experimental fronts~\cite{Grojean:2013kd,
	Jenkins:2013zja, Jenkins:2013wua, Alonso:2013hga, Elias-Miro:2013gya,
	Elias-Miro:2013mua, Alonso:2014zka, Alonso:2014rga, Trott:2014dma,
	Henning:2014wua,
	Willenbrock:2014bja,Elias-Miro:2014eia,Gupta:2014rxa,Giudice:2007fh,
Higgsbasis,Cheung:2015aba, Berthier:2015gja, Chiang:2015ura, Huo:2015exa,
Huo:2015nka, Drozd:2015kva}. This
program will only increase in intensity as
data from the second LHC run begins to accumulate, as higher statistics will
allow more precise measurements and more (i.e.~differential) observables.  The
invariant operators of the SMEFT  are relatively straightforward to construct at
low mass dimension, but the calculation becomes increasingly difficult at higher
mass dimension.  At mass dimension $d=5$ there is only 1
term~\cite{Weinberg:1979sa}, at $d=6$ there are 63~\cite{Buchmüller1986621,
Grzadkowski:2010es, Abbott:1980zj, Alonso:2014zka}, and at $d=7$ there are
20~\cite{Lehman:2014jma}. No complete set of $d=8$ operators has been presented
to our knowledge, although partial sets containing interactions  of three
neutral gauge bosons have been presented in~\cite{Gounaris:2001mw,
Degrande:2013kka}. 

Once we recast the SMEFT (or any EFT) into the formalism of a Hilbert series, we
can use powerful mathematical techniques to evaluate $\h$.  Specifically, $\h$
can be written as the Haar integration over the relevant symmetry groups of a
generating function known as the plethystic exponential
(PE)~\cite{Benvenuti:2006qr, Feng:2007ur, Hanany:2008sb, Gray:2008yu,
Hanany:2014dia}. The input to the PE is the field content of the theory,
meaning the number of flavors and the quantum numbers of the fields in question.

The PE for a (bosonic) field $\phi$ in a representation $R$ of the symmetry group
of the theory is formed by
\begin{equation}
	\text{PE}[\phi_R] = \exp\Big({\sum\limits_{r = 1}^{\infty}
	\frac{\phi^r \,\chi_{R}(z_j^r)}{r}}\Big),
	\label{eq:PE}
\end{equation}
where $\chi_R$ is the character function of the representation $R$ expanded into
monomials of $j$ complex variables, where $j$ is the rank of the group in
question. As an example, for the SM Higgs field charged under $SU(2)_w \otimes
U(1)_Y$, the entry in the above PE would be 
\begin{equation}
\phi\; \chi_R = H \Big(z + \frac 1 z\Big)u^{1/2},
\end{equation}
where $z$ is the complex variable parameterizing $SU(2)_w$ and $u$ is the
variable for $U(1)_Y$. Here and throughout, $H$ is taken to be another complex
variable of modulus $< 1$ and not a full-fledged quantum field. We will refer to
these complex numbers that represent fields as ``spurions."

For anticommuting spurions $\psi$, which represent fermionic fields, the generating
function is the fermionic plethystic exponential (PEF)~\cite{Hanany:2014dia},
defined as
\begin{equation}
	\text{PEF}[\psi_R] = \exp\Big({\sum\limits_{r = 1}^{\infty}
	\frac{(-1)^{r+1}\psi^r \,\chi_{R}(z_j^r)}{r}}\Big).
\end{equation}

The product of the PE for all bosonic spurions and the PEF for all fermionic
spurions gives the complete generating function for the theory (which we call
$\text{PE}_{\text{tot}}$) at zeroth order in derivatives. Expanding,
$\text{PE}_{\text{tot}}$ generates all possible combinations of the input
spurions. We can pick out the invariants of the symmetry group from the complete
set of combinations by integrating over the Haar measure for each individual
symmetry group included in $\text{PE}_{\text{tot}}$. For the Higgs example
above, the two individual groups are $SU(2)_w$ and $U(1)_Y$, but for the
complete SMEFT this must be extended to include $SU(3)_c$ and the Lorentz group
as well.  The wide variety of the SMEFT field content -- fermions, scalars,
gauge fields, chiral charges, vector charges, etc.~is yet another reason to use
it as a working example for illustrating Hilbert series methods.

Integrating over the Haar measure projects out the group invariants because the
characters of a compact Lie group form an orthonormal basis for class functions,
i.e.~functions of the $j$ complex variables parameterizing the group: for group
representations $R$ and $M$, $\int d\mu\, \chi_R\, \chi^*_M = \delta_{RM}$.  As
such, we can expand any function of the $j$ variables as a linear combination of
characters:
\begin{equation}
f(z) = \sum\limits_R A_R \, \chi_R(z),
\label{eq:character}
\end{equation}
where $A_R$ are coefficients that are independent of the complex number(s) used to
parameterize the group ($z$ in this case). Integrating $f(z)$ over the Haar
measure, itself a function of $z$, projects out the part of $f$ in the trivial
representation, $A_0$.  In practice, each Haar integration can be expressed as a
contour integral. (See Appendix~\ref{sec:singlephi} for an example. Explicit
expressions for the Haar measures of other simple Lie groups can be found
in~\cite{Hanany:2008sb, Lehman:2015via}.)  Taking the residues of the poles in
order to carry out this (potentially multiple) contour integral yields the
number and form of the invariants of the theory. See Ref.~\cite{Lehman:2015via}
for examples.

Our inclusion of Lorentz symmetry requires a more detailed explanation. Lorentz
symmetry must be included as soon as we include particles with spin greater than
zero, regardless of whether or not derivatives are added. The orthonormality of
the characters  just described only holds for compact Lie groups, and the
Lorentz group is not compact.  However, since we only care about operator
counting and not about dynamics, we can work in Euclidean space where the
Lorentz group is compact: $SO(4) \cong SU(2)_R \otimes SU(2)_L$.  For simplicity
we work with fundamentally left-handed fields, defined to sit in the (0, 1/2)
representation of $SU(2)_R \otimes SU(2)_L$; the hermitian conjugates are then
in the (1/2, 0) representation.  Gauge field strengths also
transform under the Lorentz group, and we work with the combinations
$X^{L,R}_{\mu\nu} = \frac 1 2 (X_{\mu\nu} \pm i \tilde X_{\mu\nu})$ (alternately
referred to as $X^-$ and $X^+$), which lie in the $(0,1)$ and $(1,0)$
representations respectively~\cite{Alonso:2014rga}.\footnote{Here
	$\tilde{X}_{\mu\nu} = \epsilon_{\mu\nu\rho\sigma}X^{\rho\sigma}/2$.} 

Applying the above discussion to the SMEFT PE, we can write $\text{PE}_{\text{tot}}$ as
a formal expansion in characters of all SM groups 
\begin{equation}
	\text{PE}_{\text{tot}}[Q, u^c, d^c,\dots, H, B^{L,R}, \dots] =
	\sum\limits_G  \Big( \sum\limits_{R_G} A_{R_G} \chi_{R_G} (z_G) \Big), 
	\label{eq:SMPE0}
\end{equation}
where the index $G$ runs over the groups $SU(2)_L$, $SU(2)_R$, $SU(3)_c$,
$SU(2)_w$, $U(1)_Y$, and the index $R$ runs over the included representations
within each group (singlet, doublet, triplet, etc.). The $A_{R_G}$ are functions
of all SM fields (as spurions) alone, while the characters are functions of the
complex numbers used to parameterize groups (the $z_G$ in Eq.~(\ref{eq:SMPE0})). The
full SMEFT Hilbert series at zeroth order in derivatives is the coefficient of
the overall (gauge and Lorentz) singlet, which we extract by integrating
the full plethystic exponential $\text{PE}_{\text{tot}}$ over all of the group volumes:
\begin{equation}
	\h_{SM} = \int \prod\limits_{G}d\mu_{G}\, \text{PE}_{\text{tot}}[Q, u^c,
	d^c,\dots,H, B^{L,R}, \dots],
	\label{eq:HIL1}
\end{equation}

Finding all of the poles and associated residues of the Hilbert series -- and
therefore implicitly finding the number of invariants at {\em all} orders of
mass dimension -- is a daunting computational task, especially for an EFT with
as many fields as the SMEFT. Instead of attacking the all-orders expression, we
will expand $\h$.  Specifically, each field (spurion) is weighted by its mass
dimension, i.e.~$Q \to \epsilon^{3/2}Q$, $H \to \epsilon\,H$, etc., and
	we expand to the desired order in $\epsilon$ (for example, take the
	coefficient of $\epsilon^6$ for the calculation of the dimension-6
	operators). After expanding and picking out the desired coefficient, all
	of the poles of the integrand are at the origin of the
	complex plane, making the residues much easier to calculate.\footnote{To
	apply these techniques in other spacetime dimensions, this mass counting
	would have to be altered along with the size of the Lorentz group.}

At a given order in mass dimension, the output of the Hilbert series is a
sequence of spurions representing an invariant combination of fields. No
information on how to actually form the invariant (i.e.~contract the group
indices) is provided, so this must be inferred from context. For concreteness,
we can carry out this reverse engineering for an example operator in the SMEFT
in Appendix~\ref{sec:indices}. 

To extend this technique to include all physical operators in an EFT, we need to
include derivatives. As mentioned earlier, derivatives bring two problems.
First, if two operators differ only by terms that reduce upon using the
equations of motion, they are redundant and only one should be included when
counting invariants.\footnote{Specifically, pieces proportional to the EOM do
	not
	contribute to on-shell matrix elements~\cite{Politzer:1980me,Georgi:1991ch, Arzt:1993gz,
	GrosseKnetter:1993td, Simma:1993ky}.}
	The second problem occurs when two
	operators differ by a total derivative and can therefore be related
	using integration by parts. This is also a redundancy and only one
	operator should be kept in the physical basis. Often, operators are related by the combined
	action of both redundancies, i.e.~IBP on one operator creates a second
	operator plus pieces that are proportional to EOM.

As an  example, if we asked for all possible invariants formed from
two derivatives and four Higgses, $\mathcal O(D^2H^4)$, we would find three
terms (the 2 in the third term is for the 2 independent ways of contracting the $SU(2)_w$
indices):
\begin{align}
(D_{\mu}H^{\dag}) (D^{\mu}H^{\dag}) \,H^2, \quad 
(D_{\mu}H )\, (D^{\mu}H) \,H^{\dag2},\quad
2\, (D_{\mu}H^{\dag}) (D^{\mu}H) \, H^{\dag}H.
\label{eq:red_ex1}
\end{align}
However, these terms are not independent since IBP on either of the first two
terms generates the third term, a total derivative, and a piece containing $\Box
H$ or $\Box H^{\dag}$. Therefore only the third term needs to be included in the
operator counting.  As a second example, consider the term at $\mathcal
O(D^4H^2)$:
\begin{align}
D^2_{\{\mu,\nu\}}H^{\dag}D^{2,\{\mu\nu\}}H,
\label{eq:red_ex2}
\end{align}
where the braces indicate the symmetric combination of indices
($D^2_{\{\mu,\nu\}} = \{D_{\mu}, D_{\nu}\}$). Under IBP, this term only
generates a total derivative and EOM terms, thus it completely vanishes from the
physical operator basis.

Recently, a method to incorporate derivatives and their redundancies was pointed
out in Ref.~\cite{berkeley} in the context of Hilbert series for scalar fields
in $(0+1)$ dimensions. The starting point of Ref.~\cite{berkeley} was to dress
each scalar field `flavor' $\phi_i$ with a series of derivatives, $\phi_i \to
\phi_i (1+ \partial_t + \partial_t^2 + \cdots)$.  Because of the low
dimensionality, the full series of both fields and derivatives could be summed,
generating expressions for invariants to all orders in $\phi_i$ and
$\partial_t$. While compact, this all-orders expression was initially plagued by
the same IBP and EOM redundancies as our Hilbert series in four spacetime
dimensions. However, Ref.~\cite{berkeley} showed these redundancies could be
removed. The key to removing the redundancies was to write the Hilbert series as
a nested sum, each power of $\phi_i$ weighted by a sum over derivatives. The
derivative portion is a polynomial in the momenta, the number of momenta
depending on the power of $\phi_i$. Each term in the derivative polynomial can
be massaged and the pieces that represent EOM or IBP redundancies can be
removed. For example, total momentum conservation could be used to swap out
combinations of momenta, and any term in the polynomial with multiple powers of
the same momentum $p^2_i$ -- the Fourier transform of the EOM term
$\partial^2_t\phi_i$ -- could be removed.  These constraints on the momentum
polynomial can be elegantly phrased in the algebraic language of rings and
ideals. After systematically cleaning up the momentum sum for each power of
$\phi_i$, then summing over the powers of $\phi_i$, the resulting Hilbert series
is free of all redundancies. For multiple flavors, the procedure is similar,
though the nested sums become more difficult to work with.\footnote{To get
	results for larger numbers of flavors, Ref.~\cite{berkeley} exploited an
underlying $SL(2,\mathbb C)$ structure in the Hilbert series, where $\phi_i$ and
its first derivative $\partial_t \phi_i$ form a doublet.}

In higher spacetime dimensions, the Lorentz group is non-trivial, thus the
transformation properties of derivatives become complicated and we cannot easily
sum over all derivatives. However, our goals are also different; the aim of
Ref.~\cite{berkeley} was the all-orders Hilbert series, while, at least for this
work, we are interested in the subset of all operators at a specified mass
dimension.  

As we will now show, we can adapt several ideas from the $(0+1)$ dimensional
method to four spacetime dimensions. We will first present the general
algorithm, explain and motivate its ingredients, and show its limitations. Along
the way we'll give several examples.

\section{Including derivatives in Hilbert series}
\label{sec:EOM}

The algorithm we have constructed to include derivatives into Hilbert series in
four spacetime dimensions and deal with their redundancies can be compactly
written as the difference of two Hilbert series, with the coefficient of each
term restricted to be positive:
\begin{equation}
	\text{max}\{\h^{\text{single}}_{EFT} - D\h^{\text{single}}_{D,\,EFT},0\}
\label{eq:himp2}
\end{equation}

Dropping the superscript for the moment, the first term is the Hilbert series
for the EFT of interest with each spurion of the theory dressed with
derivatives~\cite{Lehman:2015via}. For a single real scalar field, dressing with
derivatives means we modify the PE to
\begin{equation} 
	\text{PE}\Big[\phi \Big(1+ D\Big(\frac 1 2, \frac1 2\Big) +
D^2\Big(1,1\Big) + D^3\Big(\frac 3 2, \frac 3 2 \Big) + \cdots\Big)\Big].
\label{eq:dressed}
\end{equation}
Here, $D$ is a spurion for the derivative, and the numbers in parenthesis label
the representation under $SU(2)_R \otimes SU(2)_L$.\footnote{Here, we use 1 for
the trivial representation (0,0).} Integrated over the Haar measure, this PE
generates all invariant operators formed from $\phi$, $D_{\mu}\phi$,
$D^2_{\{\mu,\nu\}}\phi$, etc. At order $D^2$ and higher, there are multiple ways
to contract Lorentz indices, but we have only included the fully symmetric
representation in Eq.~(\ref{eq:dressed}). The other possibilities are omitted
because they are redundant by the EOM. For example, at $D^2$ there are two
possible terms: $D^2_{\{\mu, \nu\}}\phi$ and $\Box \phi$. However, in any
invariant containing a $\Box\phi$, the $\Box \phi$ can be rewritten using the
equations of motion as a term with fewer derivatives and is therefore already
included in the PE (i.e.~$\phi^2\Box\phi$ reduces to $\phi^3$). At higher
derivative orders, only the fully symmetric product of derivatives represents a
genuinely new spurion and merits addition into the PE -- all other combinations
reduce through the EOM and are therefore redundant with terms that already
exist. This process of removing the $\Box$ terms is similar in spirit to
projecting out the  $\partial^n_t, n\ge 2$ pieces of the derivative polynomial
in~\cite{berkeley}, though done here at the level of the argument of the PE.

Similar logic follows for fermions~\cite{Lehman:2015via}. Consider the EOM for the left-handed
Standard Model quark doublet $Q$:
\begin{equation}
	i \slashed{D}Q = y_u^\dagger \, u^{c \dagger}\epsilon H^* + y_d^\dagger
	\, d^{c \dagger} H,
\label{deriv_eom}
\end{equation}
which tells us that $\slashed D Q$ can be removed in favor of a combination of
fields with no derivatives. However, $\slashed D\,Q$ is just one way to contract
the Lorentz indices. Acting on a left-handed fermion with a derivative, we get
(ignoring all non-Lorentz information):
	$\left(\frac{1}{2},\, \frac{1}{2}\right) \otimes \left(0,\,
	\frac{1}{2}\right) = \left(\frac{1}{2},\, 0\right)
	\oplus \left(\frac{1}{2},\, 1\right)$.
The EOM in Eq.~(\ref{deriv_eom}) involves the $(\frac 1 2, 0)$ representation
only. Following the same prescription we used for our toy scalar theory, we can
incorporate the EOM for a fermion field $Q$ (for example) by including an
additional spurion for the $(\frac 1 2, 1)$ part of $D_{\mu}Q$, but omitting the
$(\frac 1 2, 0)$ spurion.  As with scalars, only the completely symmetric
combinations of higher derivative terms generate new invariants, i.e.~
$D^2_{\{\mu\nu\}}Q \sim (1,\frac 3 2)$. Analogously, for derivatives of the
field strength tensors $D_{\lambda}X^{\pm}_{\mu\nu}$, the $(\frac 1 2, \frac 1
2)$ representation should be omitted, while the symmetric combination is kept
(the $(\frac 1 2, \frac 3 2)$ or $(\frac 3 2, \frac 1 2)$).

One may worry that once we consider a gauge theory, the ordinary derivatives
all become covariant derivatives (we have been sloppy in the above discussion
about distinguishing between the two) and it is no longer true that derivatives
commute. This is not an issue; $[D_{\mu}, D_{\nu}]$ has the same quantum numbers
as $X^{L,R}_{\mu\nu}$, so the antisymmetric combinations of derivatives acting
on a field are redundant with the field strength times that field~\cite{Grzadkowski:2010es}. Combined with
the redundancy pointed out above, we can summarize the action of two derivatives
on the fermion field $Q$:
\begin{equation}
D^2 \times Q = D^2_{\{\mu\nu\}}Q + X^L_{\mu\nu}Q + X^R_{\mu\nu}Q + \Box Q,
\end{equation}
where $X^{L,R}_{\mu\nu}$ stands for the field strength of any of the SM gauge groups.
The last three terms are already present in the PE with fewer derivatives and
therefore are redundant, so only $D^2_{\{\mu\nu\}}Q$ needs to be added. Higher
derivatives and fields with other spins work the same way.

We will call the combination of the derivative-dressed PE for each spurion
(scalars $\phi$, fermions $\psi$, or gauge fields $X$) $\text{PE}_{EFT}[\phi,\psi, X]$.
One important feature of $\text{PE}_{EFT}$ above is that the same spurion $D$
appears wherever there is a derivative, i.e.~$D \phi$ is really the product of
two separate spurions and $D \phi$ and $D \psi$ involve the same $D$. This will
be important for keeping track of the total number of derivatives in an
invariant, which we need to do to in order to resolve the integration by parts
redundancy. Another important feature is that the derivative dressing for each
given spurion is technically an infinite series. While it is interesting to
think about this infinite sum of derivatives, and we present some thoughts in
Appendix~\ref{sec:singlephi}, we will see shortly that Eq.~(\ref{eq:himp2})
fails at high derivative order so for our purposes  we only need to keep
$\mathcal O(D)$ and $\mathcal O(D^2)$.

Returning to Eq.~(\ref{eq:himp2}), we now discuss the second term
$\h_{D,\,EFT}$. This term contains all terms generated by the dressed PE that
lie in the $(\frac 1 2, \frac 1 2)$ (four-vector) representation of the Lorentz
group but are otherwise invariant
\begin{equation}
 \h_{D\, , EFT} = \int
d\mu_{SU(2)_L}d\mu_{SU(2)_R}\,\Big(\frac 1 2, \frac 1 2\Big)\,
\text{PE}_{EFT}[\phi, \psi, X].
\end{equation}
Some examples are $\phi\, D_{\mu}\,\phi,\, D^2_{\{\mu,\nu\}}\phi\, D^{\nu}\phi\,
\phi^2$, etc. We project out this subset using the same character orthogonality
as in  Eq. (\ref{eq:HIL1}). The argument of the integration is the same,
$\text{PE}_{EFT}$; the only change is that we include a factor of the $(\frac 1 2,
\frac 1 2)$ character function so that the objects in the four-vector
representation are projected out.

Having identified $ \h_{EFT}$ and $ \h_{D,\, EFT}$, the logic behind subtracting
the two in Eq.~(\ref{eq:himp2}) is that any term with $2n$ derivatives can be formed
by acting on a term with $2n-1$ derivatives with one additional derivative.
Schematically, for scalar fields:
\begin{equation}
\mathcal O(D^{2n}\phi^m) = D\times \mathcal O(D^{2n-1}\phi^m),
\end{equation}
with higher spin fields behaving similarly.  Now, lets assume that there are $j$
operators at $ O(D^{2n}\phi^m)$ (taking scalars for simplicity) and $k$
operators at $\mathcal O(D^{2n-1}\phi^m)$. As all of
the $k$ $\mathcal O(D^{2n}\phi^m)$ terms must be generated from the same
$\mathcal O(D^{2n-1}\phi^m)$ set with $j$ elements, the number of independent
$\mathcal O(D^{2n}\phi^m)$ is the difference: 
\begin{equation}
(\#\, \text{invariants}) - (\#\, \text{relations}) = k - j.
\label{eq:IBP}
\end{equation}
Any terms that can be connected through a total derivative are removed by this
subtraction, thus we are left with the number of invariants after taking
integration by parts redundancies into account. The extra power of the spurion $D$
accompanying the $ \h_{D,\, EFT}$ is necessary to ensure that the dimension and
derivative counting matches in the two terms.

For low numbers of derivatives, this subtraction works perfectly. For example,
lets apply this algorithm to a theory of a single scalar field at
$\mathcal{O}(D^2\phi^m)$. There is only one term in the Hilbert series $\h_\phi$
at $\mathcal{O}(D^2\phi^m)$, namely $(D_{\mu}\phi)\, (D^{\mu}\phi)\,
\phi^{m-2}$. However
\begin{equation}
(D_{\mu}\phi)\, (D^{\mu}\phi)\, \phi^{m-2} \sim (D_{\mu}\phi)\, D^{\mu}(\phi^{m-1})
\xrightarrow[\text{IBP}]{}  \phi^{m-1}\Box \phi \rightarrow EOM,
\label{eq:d2phim}
\end{equation}
so all $\mathcal{O}(D^2\phi^m)$ terms are actually redundant, via EOM and
dropping total derivatives, with terms that have fewer derivatives (and are
already in the PE).\footnote{Throughout this paper, we are only concerned with
	the number of operators. The coefficient (including sign) in front of
	the operator is unimportant and will be neglected.} There is only one
	$\mathcal{O}(D\phi^m)$ term, namely $(D_{\mu} \phi)\,\phi^{m-1}$,
	generated by $\h_{D\phi}$, so the prescription of Eq.~(\ref{eq:himp2})
	for the improved Hilbert series correctly gives $0$ terms. 

Terms at $\mathcal O(D^4\phi^m)$ are also correctly counted by this method,
however things go awry when we go to even higher derivatives. Focusing on terms
with four $\phi$ fields, one can show that there are 6 operators at
$\mathcal{O}(D^6\phi^4)$ and 6 at $\mathcal{O}(D^5\phi^4)$. This would indicate
that no invariants are left after IBP. However, if we more closely examine the relations generated by
the 6 $\mathcal{O}(D^5 \phi^4)$ operators, we see that only 5 are linearly
independent. The simplest way to see this is to cast $D_{\mu}
\,\mathcal{O}(D^5\phi^4) = 0$ as a matrix times a vector of the
$\mathcal{O}(D^6\phi^4)$  invariants, then determine the rank of the matrix.
Subtracting the number of independent constraints from the number of operators,
we find one operator at $\mathcal{O}(D^6\phi^4)$, consistent with direct
computation. The same thing happens at $\mathcal{O}(D^8\phi^4)$; there are 11
invariants at $\mathcal{O}(D^6\phi^4)$, and 15 at $\mathcal{O}(D^7\phi^4)$,
though only 10 of the 15 resulting relations are linearly independent. 

Comparing these scalar examples, we see that the subtraction of
Eq.~(\ref{eq:himp2}) correctly removes
IBP redundancy only when the constraint equations (i.e $D\times
\mathcal{O}(D^{2n-1}\phi^4) = 0$ for the case of 4 fields) are linearly independent. The
independence of the constraint equations is linked to the number of ways to
partition the derivatives among the fields and the number of Lorentz invariants
per partition. Using a notation $D^i\phi\, D^j\phi\, D^k\phi\, D^m\phi =
(i,j,k,m)$ and ranking powers of $D$ in descending order, we can translate any
given field content (number of derivatives $D$ and number of fields $\phi$) into
a set of integer partitions representing different operators that can be formed
from that field content: $D^4\phi^4\, \to (2,2,0,0),\, (2,1,1,0),\, (1,1,1,1)$;
$D^6\phi^4 \to$ $(3,3,0,0)$, $(3,2,1,0)$, $(3,1,1,1)$, $(2,2,2,0)$, $(2,2,1,1)$,
etc. For a some partitions, such as $(2,2,0,0)$, there is only one way to contract Lorentz
indices, namely $D_{\{\mu,\nu\}}\phi\, D^{\{\mu,\nu\}}\phi\, \phi^2$; however at higher
orders there are sometimes multiple possibilities. For example, the partition $(2,2,1,1)$
for $D^6 \phi^4$ has two Lorentz contractions:
\begin{align}
D^2_{\{\mu,\nu\}}\phi\, D^{2,\{\mu,\nu\}}\phi\, D_{\rho}\phi\,
D^{\rho}\phi \quad \text{and} \quad
D^2_{\{\mu,\nu\}}\phi\, D^{2,\{\nu,\rho\}}\phi\,
D^{\mu}\phi\, D_{\rho}\phi .
\end{align}
If there is only one contraction per partition, Eq.~(\ref{eq:himp2}) simply
counts the number of partitions of $2n$ derivatives among $m$ fields minus the
number of partitions of $2n-1$ derivatives on the same number of fields. Each of
the partitions of $2n-1$ derivatives gives a single independent constraint
equation. For example, applying a derivative to a term $\mathcal O(\phi^{n+4})$
where four of the $\phi$ fields carry at least one derivative each yields the
partitions:
\begin{align}
 D \otimes (D^i\phi\, D^j\phi\, D^k\phi\, D^m\phi\, \phi^n)& =   (i + 1,j,k,m,0,\cdots) + (i,j+1,k,m,0,\cdots) + \nonumber \\
& \quad\quad  (i,j,k+1,m,0, \cdots) + (i,j,k,m+1,0,\cdots) + n\,
(i,j,k,m,1,0,\cdots). \nonumber
\end{align}
If there is only one contraction per partition, then since the partitions are
independent, so are the resulting constraint equations. On the other hand, if
multiple contractions are possible, a given $2n-1$
derivative partition may yield multiple constraint equations and the
independence of the constraints breaks down.\footnote{As further proof that the
	existence of multiple Lorentz contractions per derivative partition is
	what spoils the algorithm, we can apply Eq.~(\ref{eq:himp2}) to a $0+1$
	dimensional theory. In the language we have been working with, the
	Hilbert series is $\text{PE}[\phi(1 + D)]$ (for a single field), and the analog
	of Eq.~(\ref{eq:himp2}) is $\text{max}\{(1-D)\text{PE}[\phi(1+D)], 0\}$. There
		are no Lorentz contractions to worry about in 0+1 dimensions,
		and $\text{max}\{(1-D)\text{PE}[\phi(1+D)], 0\}$ correctly captures {\em
		all} operators of the Hilbert series as presented in Eq.
		(3.9-3.12) of Ref.~\cite{berkeley}. }

Clearly, one would like an implementation of derivatives in Hilbert series that
can overcome the complication of multiple Lorentz contractions. It is possible
to modify the subtraction by adding more terms, such that Eq.~(\ref{eq:himp2})
is correct for certain higher order terms, however we have not found a
modification that works consistently for multiple types of fields or corrects
all higher order terms. The pursuit of the full implementation of derivatives is
enticing and an active area of research (and it is possible that the full
implementation of derivatives will look nothing like this subtraction), but
beyond the scope of this paper. Instead, we will carry on with
Eq.~(\ref{eq:himp2}), adding the superscript ``single" to both pieces to
indicate that Eq.~(\ref{eq:himp2}) only applies when there is a single Lorentz
contraction per derivative partition. While restricted, Eq.~(\ref{eq:himp2}) is
still sufficient to capture a large set of low-derivative operators. To get a
better idea of where Eq.~(\ref{eq:himp2}) works and where it fails, we work
through some examples. However, before proceeding, we emphasize that the Hilbert
series of the derivative-dressed PE ($\h_{EFT}$ in Eq.~(\ref{eq:himp2})) does
generate the full set of invariants of fields and their derivatives while
correctly accounting for EOM redundancy. As such, even in cases where
Eq.~(\ref{eq:himp2}) fails, $\h_{EFT}$ alone will give an {\em upper bound} on
the number of operators at a given mass dimension. Furthermore, one can always
take the output of $\h_{D, EFT}$ for a given set of fields, apply another
derivative and (albeit, tediously) compute the rank of the resulting constraint
matrix by hand. So, the spirit of Eq.~(\ref{eq:himp2}) is certainly true; what
fails is the automatic calculation of independent constraints. 

\subsection{Example: Complex scalar}
\label{sec:phistar}

For a single real scalar field we saw that Eq.~(\ref{eq:himp2}) becomes
untrustworthy once we go to $> 6$ derivatives and $> 4$ fields, or operators of
dimension ten or higher. Moving to a theory of a single complex scalar
field, all invariant operators must have the field content
$D^{2n}\phi^m\phi^{*m}$.  Consider the operators with $m = 2$.  As there are two
distinct fields, we have to adapt the partition notation to $(D^i\phi D^j\phi
D^k\phi^* D^m\phi^*) \equiv (i,j;k,m)$, where numbers of derivatives acting on
$\phi$ and $\phi^*$ are separately listed in descending order and the division
between $\phi$ and $\phi^*$ is indicated by the semicolon. This notation
generalizes straightforwardly when there are even more fields present, i.e. $D^i\phi_1\,D^j\phi_2\, D^k\phi_3\, \phi_3 = (i;j;k,0)$.

At $\mathcal O(D^2)$ there are three partitions: $(1,0; 1,0), (1,1; 0, 0)$ and
$(0, 0; 1,1)$; each with a single Lorentz contraction. At four derivatives, the
8 partitions are $(2,2; 0, 0)$, $(0,0 ; 2, 2)$, $(2,1; 1, 0)$, $(1,0; 2, 1)$,
$(2,0; 1, 1)$, $(1,1; 2, 0)$, $(2,0; 2,0)$ and $(1,1; 1,1)$. The last partition
has two possible Lorentz contractions ($D_{\mu}\phi\, D^{\mu}\phi\,
D_{\nu}\phi^*\, D^{\nu}\phi^*$ and $D_{\mu}\phi\, D^{\mu}\phi^*\,D_{\nu}\phi\,
D^{\nu}\phi^*$), so there are 9 operators before accounting for IBP.  As
expected, Eq.~(\ref{eq:himp2}) miscounts in this situation, predicting 1
invariant rather than the 2 independent invariants found the brute force way by
calculating the rank of the constraint matrix $M$ (there are 8 constraint
equations from IBP but $\text{rank}(M) = 7$). This example reinforces the notion
that the subtraction in Eq.~(\ref{eq:himp2}) is limited by the presence of
multiple Lorentz contractions and not at some fixed derivative order. 

\subsection{Example: Dimension-6 SMEFT}
\label{sec:smeft}

Once we admit higher spin fields or internal symmetries, there are countless
higher dimensional, high derivative operators one could look at to see whether
they require multiple Lorentz contractions for a given derivative partitioning.
Rather than march through possibilities, let us focus on operator types that
appear in the SMEFT expansion, starting with dimension-6. While only a subset of
all possibilities, the SMEFT operators will test how Eq.~(\ref{eq:himp2})
performs on a wide array of spin and gauge symmetry structures. We will roughly
follow the formalism of Ref.~\cite{Grzadkowski:2010es}: $\psi$ will stand for
any fermion of either chirality, $X$ for any field strength (L or R), but we
will use $H$ for a Higgs (both the field and its conjugate).

We only need to check  Eq.~(\ref{eq:himp2}) against the SMEFT operators
containing derivatives. Non-derivative terms will be correctly captured
following the procedure in Ref.~\cite{Lehman:2015via} and recapped in the
introduction.

At dimension 6, the highest number of derivatives that can appear is 4: a term
of $\mathcal{O} (D^4 H^{\dag}\,H)$. As there are two fields, there is only one
partition of the derivatives, and only one Lorentz contraction for that
partition.  

Continuing to lower numbers of derivatives following~\cite{Grzadkowski:2010es},
we next have $\mathcal O(D^3\psi^2)$, where $\psi$ is either a left or
right-handed fermion. Remembering that the derivative spurions in the PE for
fermions are $D_{\mu}\psi_L \sim (\frac 1 2, 1), D_{\mu}\psi_R \sim (1, \frac 1
2)$, we can see that there is no way to partition the three derivatives among
any pair of fermions that yields a Lorentz invariant,  regardless of their
chirality (even forgetting other quantum numbers). In other words, this term is
redundant by EOM alone and would not even be generated by the derivative dressed
SMEFT PE.\footnote{Terms of $\mathcal O(D^4X)$ do not appear for the same
reason.}

At two derivatives there are four possible classes: $\mathcal O(D^2\psi^2H)$,
$\mathcal O(D^2\phi^2 X)$, $\mathcal O(D^2\, X^2)$, and $\mathcal O(D^2\,
H^2H^{2\dag})$. 
\begin{itemize}
\item $\mathcal O(D^2\psi^2H)$. In this class, invariants can only be formed if
	both fermions have the same chirality and either both of the $\psi$
	(which need not be the same) are acted on by  a derivative, or one
	derivative acts on $\psi$ and one on $H$. In terms of partitions
	$D^i\psi_a; D^j\, \psi_b; D^k H = (i;j;k)$, these possibilities are
	$(1;1;0)$, $(1;0;1)$ and $(0;1;1)$, and in all cases there is only
	Lorentz one contraction per partition.\footnote{As listed, the partitions
		are for three different fields, i.e $e_c\, L\, H$. If both
		fermions were the same -- a moot case for the SMEFT given
	$SU(2)_w$ and $U(1)_Y$ quantum numbers -- we would indicate partitions
	without the first semicolon.}
\item  $\mathcal O(D^2H^{\dag}H X)$: Following our derivative dressing,
	derivatives of field strengths are only generated by the PE in (larger)
	Lorentz representations which have no way of forming singlet, so we
	only have to consider derivatives on the scalars. There is only one
	partition, $(1;1;0)$ in an obvious extension of the partition language
	we've been using, and only one contraction for that partition.
\item $\mathcal O(D^2\, X^2)$: For $(X^L)^2, (X^R)^2$ there is a single
	invariant partition $(1,1)$, with one contraction. For $X^L\,X^R$ there
	are no Lorentz invariants. 
\item $\mathcal O(D^2\, H^2H^{2\dag})$. Both derivatives cannot act on a single
	field and yield a Lorentz invariant, but all other partitions are
	allowed: $(1,1; 0,0)$, $(0,0; 1,1)$, and $(1,0; 1,0)$. For the first
	two, there is only one Lorentz contraction and only one $SU(2)_w$
	contraction; Bose symmetry forces the pairs of spurions, both the pair
	with derivatives and the pair without, to sit in the triplet
	representation of $SU(2)_w$. The last partition also has only one
	Lorentz contraction, but has two $SU(2)_w$ contractions since no two
	fields in the operator are identical.
\end{itemize}

Finally, the two classes with one derivative are $\mathcal O(D\, \psi^2\, X)$ and
$\mathcal O(D\,\psi^2 H^{\dag}H)$.

\begin{itemize}
\item $\mathcal O(D\, \psi^2\, X)$: To be Lorentz invariant, the derivative must
	act on one of the fermions, and both fermions must have the same
	chirality. The partitions are $(1;0;0)$ (and $(0;1;0)$ if the $\psi$
	represent two different fields). One one contraction is possible.
\item $\mathcal O(D\,\psi^2 H^{\dag}H)$: The derivative must act on one of the
	scalars, so the allowed partitions are $(0;0;1;0)$ and $(0;0;0;1)$.
	Depending on the $SU(2)_w$ quantum numbers of the fermions (which must
	have opposite chirality), there may be more than one $SU(2)_w$
	invariant, but each partition only allows a single Lorentz invariant.
\end{itemize}

Having exhausted the list of dimension-6 SMEFT operators, we find that all
operators have the form amenable to Eq.~(\ref{eq:himp2}), namely that each
partition of derivatives (if any) among the fields in an operator only admits a
single Lorentz contraction. Notice we haven't yet tested for the number of
actual invariants for a given operator class; this will be done by
Eq.~(\ref{eq:himp2}). So far we have simply tested whether these classes of
operators fit the restrictions on whether Eq.~(\ref{eq:himp2}) can be used. To
apply Eq.~(\ref{eq:himp2}), we form the derivative-dressed SM PE:
\begin{align}
\text{PE}^{\text{SM}}_{\text{tot}} = &\text{PE}\Big[ H\Big(0,0; 0,\frac 1 2,
	\frac 1 2\Big) + D
	H\Big(\frac 1 2,\frac 1 2; 0,\frac 1 2, \frac 1 2\Big) + D^2 H\Big(1,1;
	0, \frac 1 2, \frac 1 2\Big) + \cdots  \nonumber \\
& + B^R \Big(1,0; 0,0,0\Big) + D B^R\Big(\frac 3 2, \frac 1 2; 0,0,0\Big) +
\cdots + B^R \to W^R, G^R + c.c. \Big] \times  \nonumber \\
& \text{PEF}\Big[ Q \Big(0,\frac 1 2; 3, 2, \frac 1 6\Big) + D Q \Big(\frac 1 2,1; 3,
2, \frac 1 6\Big) + \cdots + Q \to u_c, d_c, L, e_c + c.c.\Big], 
\label{eq:SMPE}
\end{align}
Here the numbers
in parenthesis indicate the representation/quantum numbers under $(SU(2)_R$,
$SU(2)_L$; $SU(3)_c$, $SU(2)_w$, $U(1)_Y)$ respectively (not to be confused with a derivative partition). Also, to clarify, the
notation $c.c.$ means adding in a new spurion for the conjugate of each field.
For example, we need to add the spurion $H^\dagger$ for the Higgs conjugate
field, along with the appropriate quantum numbers. Likewise, for fermions we need
to add a spurion $\psi^\dagger$ for each fermion field $\psi$.
As we are only interested in operators with less than four derivatives we
can truncate each field's derivative dressing at $\mathcal O(D^2)$. Plugging
$\text{PE}^{\text{SM}}_{\text{tot}}$ into Eq.~(\ref{eq:himp2}), and performing
the Haar integration, we find the following derivative terms at mass dimension
6. 
\begin{equation}
\begin{gathered}
D (d^{\dag}_c\,d_c\,H^{\dag}H), \enskip
D (e^{\dag}_c\,e_c\,H^{\dag}H), \enskip
2\, D (L^{\dag}\,L\,H^{\dag}H), \enskip
D (u^{\dag}_c\,u_c\,H^{\dag}H), \\
2\, D (Q^{\dag}\,Q\,H^{\dag}H), \enskip
D (H^2\,d^{\dag}_c\, u_c) + \hc, \enskip
2\,D^2 (H^{\dag}H)^2 .
\end{gathered}
\end{equation}
Notice that the derivatives appear as just another spurion in the sequence in
the Hilbert series output. No information on where the derivative acts is
provided, and this has to be figured out just like the other indices (see
Appendix~\ref{sec:indices}). 

This set of derivative terms matches exactly with
Ref.~\cite{Grzadkowski:2010es}. Combined with the zero derivative, baryon number
conserving terms that we already know will match from the analysis
in~\cite{Lehman:2015via}, we reproduce the full set of 59 dimension-6
operators.\footnote{Since we use $X^L, X^R$ for gauge field strengths, our
	purely bosonic operators are not self-conjugate, while they are in the
	notation of \cite{Grzadkowski:2010es}.  To compare with
	\cite{Grzadkowski:2010es}, we need to include Hermitian conjugates for
all purely boson operators.}
With one flavor of fermions, we find 4 baryon-number violating operators, all
with zero derivatives. Adding these to the previous set  gives a total of 63
operators, matching the total given in \cite{Grzadkowski:2010es} aside from a
single baryon-number-violating operator that vanishes in the single-flavor
($N_f=1$) case that we are considering.

As a further check, we can repeat the calculation including multiple families of
fermions, sending $Q \to N_f\,Q$, etc.~in the full $\text{PE}^{\text{SM}}_{\text{tot}}$ (Eq.~(\ref{eq:SMPE})). The
total number of operator coefficients for the baryon-number-conserving
dimension-6 operators for various $N_f$ have been calculated previously in
Ref.~\cite{Alonso:2013hga}, and we find these numbers are reproduced exactly by
Eq.~(\ref{eq:himp2}) (specifically 76 coefficients for $N_f = 1$ and 2499
coefficients for $N_f = 3$).

The success of Eq.~(\ref{eq:himp2}) in reproducing the dimension 6 SMEFT teaches
us several things. First, it provides us with several examples that
Eq.~(\ref{eq:himp2}) applies to operators containing fermions and field
strengths as well, provided they lie within the restricted class of operators
with a single Lorentz contraction per derivative partition. Second, we see that
adding global or gauge symmetry to the theory does not immediately disrupt the
subtraction setup -- multiple internal symmetry contractions on an operator do not
invalidate it from Eq.~(\ref{eq:himp2}), only multiple Lorentz contractions.
Finally, we see that Eq.~(\ref{eq:himp2}) removes IBP redundancies even when
operators contain fields with different gauge transformation properties. One may
have worried that the gauge field pieces that are present in, i.e $D_{\mu}Q$ but
not $D_{\mu}H$ (or, even more generally pieces in $D_{\mu}$ but not
$\partial_{\mu}$) get subtracted off. However, the terms in $\h_{D, EFT}$ are
only terms in the $(1/2, 1/2)$ representation of the Lorentz group and trivial
representation of all other groups: $(1/2, 1/2; 0\cdots)$, where the dots
indicate no charge under any other quantum numbers, either local or global.
Whenever $D_{\mu}$ contains non-abelian interactions, the non-abelian gauge
field pieces sit in the adjoint representation of their respective gauge group
(in addition to their Lorentz group representation) and not in $(1/2, 1/2;
0\cdots)$ alone. In other words, the subtraction in $\h_{D,EFT}$ appears to
remove redundancies from shifting the $\partial_{\mu}$ piece among fields, with
the extra pieces required to covariantize (or un-covariantize) the derivative
simply dragged along to maintain gauge invariance.

The set of dimension-6 operators given in \cite{Grzadkowski:2010es} is not a
unique representation. Operators can be related to combinations of other
operators through the equations of motion, as we have seen, and also through
general field redefinitions. While the number of parameters needed to describe
the effective field theory at a given mass dimension cannot change, by picking a
particular set, be it the set in Ref.~\cite{Higgsbasis, Hagiwara:1993ck,
	Grzadkowski:2010es,
Gupta:2014rxa,Giudice:2007fh}, we are choosing an operator basis.\footnote{By
this we mean the basis for a fixed mass dimension.} 
We have landed in the basis used by \cite{Grzadkowski:2010es} due to our
treatment of the equations of motion, whereby we removed terms whenever possible
in favor of operators with fewer derivatives.  While the choice of basis cannot
matter when calculating a physical quantity (provided one works with a complete
basis), picking the appropriate basis can certainly make a calculation simpler
or make the mapping from a UV theory to the set of effective operators
easier~\cite{Gupta:2014rxa, Willenbrock:2014bja,Higgsbasis}.

\subsection{Example: Dimension-7 SMEFT}

A further cross-check on Eq.~(\ref{eq:himp2}) is provided by using it to
calculate the dimension-7 SMEFT operators. Doing this calculation for $N_f = 1$
results in the following 6 operators containing one and two derivatives (all
plus $\hc$):
\begin{equation}
	D (e_c^\dag d_c^3), \enskip
	D (L e_c^\dag H^3), \enskip
	D (L Q^\dag d_c^2), \enskip
	D (L^2 d_c u_c^\dag), \enskip
	2\, (D^2 L^2 H^2).
\end{equation}
Combining these operators with the zero-derivative operators, we find 15
operators, matching the results of \cite{Lehman:2014jma}, up to operators that
vanish in the flavor-diagonal ($N_f = 1$) case.\footnote{An explicit calculation was done
for $N_f = 3$ using the Hilbert series, and these non-flavor-diagonal operators
are indeed produced.}
This success tell us all operators containing derivatives at dimension-7 admit
only one Lorentz contraction per derivative partition. As this is non-trivial,
we can go through the possible derivative operators in the same fashion we did
for the dimension-6 operators. Following Ref.~\cite{Lehman:2014jma}, and using
the same notation as in the previous section, there are two possible operator
classes with one derivative: 
\begin{itemize}
\item $\mathcal O(D\, \psi^2\, H^3)$: To form a Lorentz invariant, the fermions
	must have opposite chirality and the derivative be applied to one of the
	Higgses. There may be multiple $SU(2)_w$ contractions, but only one
	Lorentz contraction is possible.
\item $\mathcal O(D\,\psi^4)$: A Lorentz singlet only arises when three fermions have
	one chirality, one of which carries the derivative, and the fourth
	fermion has the opposite chirality, i.e. $\psi_R\, D\psi_L\, \psi^2_L$.
	There may be several internal symmetry configurations that are
	permitted, or it may be the case that internal symmetry considerations
	only allow the derivative on a subset of the same-chirality fermions. In
	either case, there is only one Lorentz contraction.
\end{itemize}
At two derivatives, there is only a single class:
\begin{itemize}
\item $\mathcal O(D^2\, \psi^2\, H^2)$: Lorentz singlets can be formed if both
	$\psi$ (same chirality) are acted on by a derivative, both $H$, or one
	$\psi$ and one $H$: partitions of $(1,1; 0,0)$, $(0,0; 1,1)$, or
	$(1,0;1,0)$ respectively. Partitions $(1,1; 0,0)$ and $(1,0;1,0)$ are
	similar to cases considered earlier, and there is one Lorentz
	contraction in each case. The $(0,0; 1,1)$ partition is different as the
	Lorentz and $SU(2)_w$ representations are linked by Bose/Fermi
	statistics. Specifically, the fermion pair, which must be a lepton
	doublet $L^2$ following the logic in~\cite{Lehman:2014jma} can either
	reside in the singlet $(0,0)$ or triplet $(0,1)$ representation of the
	Lorentz group, and each of those can be contracted with $(DH)^2$.
	However, the Lorentz singlet $L^2$ must be an $SU(2)_w$ triplet, while
	the Lorentz triplet combination must be an $SU(2)_w$ singlet. Thus,
	while there are multiple Lorentz contractions for the $(0,0; 1,1)$
	derivative partition, there is only {\em one contraction per partition
	for a given assignment of internal symmetry charges}.
\end{itemize}

So, while Eq.~(\ref{eq:himp2}) holds for all dimension-7 operators, we have
learned that the internal quantum numbers can be important. This means that,
when considering the SMEFT at higher mass dimension or when looking at EFT
outside of the SM, we cannot determine the viability of Eq.~(\ref{eq:himp2})
based on the Lorentz properties of the field in the operators alone.

Finally, as the calculation of the number of dimension-7 operators for arbitrary
$N_f$ has not yet been done in the literature, the results of this Hilbert
series computation are shown in Appendix \ref{dim_7_3}.


\section{SMEFT: dimension-8 operators}
\label{sec:dim8}

We can now use the prescription of Eq.~(\ref{eq:himp2}) to study SMEFT
dimension-8 operators for $N_f = 1$. We have learned from previous examples,
Eq.~(\ref{eq:himp2}) correctly subtracts off IBP redundancies only when there is
a single Lorentz contraction for a given partition of derivatives and assignment
of internal symmetry charges. As we will see, these restrictions prevent us from
automatically generating the full set of dimension 8 SMEFT operators. However,
we can generate a larger subset than has been shown previously, and set up a
prescription for filling in the remaining operator classes.

To begin with, as no complete list of dimension-8 SMEFT operators without
derivatives have been shown, we list the terms at $\mathcal O(D^0)$. These can be obtained following the recipe in the introduction using $\text{PE}^{\text{SM}}_{\text{tot}} $ defined in Eq.~(\ref{eq:SMPE}). Readers
interested only in which derivative terms are allowed and which are not should
skip to Sec.~\ref{sec:dim8derivs}. 

\subsection{Derivative-free operators}

Following Ref.~\cite{Grzadkowski:2010es, Lehman:2014jma}, we can group terms
into classes governed by number of scalars, fermions, and field strengths. We
will continue to use $H$ to represent Higgses (field or conjugate), $\psi$ for
any fermion, and $X$ for any field strength. At dimension-8 and no derivatives,
there are 9 operator classes: $\mathcal O(H^8)$, $\mathcal O(\psi^2\, H^5)$,
$\mathcal O(\psi^4\, H^2)$, $\mathcal O(\psi^4\, X)$, $\mathcal O(\psi^2\, H\,
X^2)$, $\mathcal O(\psi^2\, H^3\, X)$, $\mathcal O(H^4\,X^2)$, $\mathcal O(H^2\,
X^3)$ and $\mathcal O(X^4)$\footnote{While operators $\mathcal O(H^6\, X)$ have
dimension 8, there is no way to make a Lorentz invariant without derivatives,
thus there are no operators of this type at $\mathcal O(D^0)$.}. \\

\noindent \underline{Class  $\mathcal O(H^8)$}:  At 8 Higgs fields (counting both $H$ and $H^\dag$ as
Higgses) there is only one term
\begin{align}
(H^{\dag}H)^4.
\end{align}

\noindent \underline{Class $\mathcal O(\psi^2\, H^5)$}: This class contains 3 operators (all plus $\hc$)
\begin{equation}
(L\,e_c\, H^{\dag})(H^{\dag}H)^2, \enskip
(Q\, d_c\, H^{\dag})(H^{\dag}H)^2, \enskip
(Q\, u_c\, H)(H^{\dag}H)^2.  
\end{equation}

\noindent \underline{Class $\mathcal O(\psi^4\, H^2)$}: We can divide this class further depending on the chirality of the fermions. There are 7 operators with two Higgses
and only $SU(2)_w$ singlet fermions:
\begin{equation}
\begin{gathered}
(e^{\dag}_c\, e_c)^2\,(H^{\dag}H), \enskip
(d^{\dag}_c\, d_c)^2\,(H^{\dag}H), \enskip
(u^{\dag}_c\, u_c)^2\,(H^{\dag}H), \\
(d^{\dag}_c\, d_c)(e^{\dag}_c\, e_c)\,(H^{\dag}H), \enskip
(u^{\dag}_c\, u_c)(e^{\dag}_c\, e_c)\,(H^{\dag}H), \enskip
2\,(u^{\dag}_c\, u_c)(d^{\dag}_c\, d_c)\,(H^{\dag}H).
\end{gathered}
\end{equation}
and 10 operators with only $SU(2)_w$ doublet fermions. Both the all-doublet and all-singlet operators sets involve self-Hermitian operators
\begin{equation}
\begin{gathered}
2\,(L^{\dag}L)^2\, (H^{\dag}H), \enskip
3\,(Q^{\dag}Q)^2\, (H^{\dag}H), \enskip
5\,(L^{\dag}L)(Q^{\dag}Q)\, (H^{\dag}H).
\end{gathered}
\end{equation}
Operators involving both $SU(2)_w$ singlet and doublet fermions can be either self-Hermitian or not. There are 16 self-Hermitian operators:
\begin{equation}
\begin{gathered}
2\,(L^{\dag}\,L)(d^{\dag}_c\, d_c)(H^{\dag}H), \enskip
2\,(L^{\dag}\,L)(u^{\dag}_c\, u_c)(H^{\dag}H), \enskip
2\,(L^{\dag}\,L)(e^{\dag}_c\, e_c)(H^{\dag}H), \\
2\,(Q^{\dag}\,Q)(e^{\dag}_c\, e_c)(H^{\dag}H), \enskip
4\,(Q^{\dag}\,Q)(d^{\dag}_c\, d_c)(H^{\dag}H), \enskip
4\,(Q^{\dag}\,Q)(u^{\dag}_c\, u_c)(H^{\dag}H),
\end{gathered}
\end{equation}
and 21 operators with Hermitian conjugates:
\begin{equation}
\begin{gathered}
e^2_c\, L^2\, (H^{\dag})^2, \enskip
2\, d^2_c\, Q^2\, (H^{\dag})^2, \enskip
2\, u^2_c\, Q^2\, H^2, \enskip
2\, (d_c\, Q)(e_c\, L)\, (H^{\dag})^2, \\
2\, (d_c\, Q)(e^{\dag}_c\, L^{\dag})(H^{\dag}H), \enskip
(d^{\dag}_c\, u_c)(L^{\dag}\, L)H^2, \enskip
4\,(u_c\, Q)(e_c L)(H^{\dag}H), \\
(u_c\, Q)(e^{\dag}_c\, L^{\dag})H^2, \enskip
4\,(u_c\, Q)(d_c\, Q)(H^{\dag}H), \enskip
2\,(u_c\, Q)(d^{\dag}_c\, Q^{\dag}) H^2.
\end{gathered}
\end{equation}
Finally, there are 6 baryon-number-violating operators (all plus $\hc$):
\begin{equation}
\begin{gathered}
2\,L\,Q^3\, (H^{\dag}H), \enskip
2\,(u_c\,d_c)(Q^{\dag}\,L^{\dag})(H^{\dag}H), \enskip
(e_c u_c)(Q^{\dag})^2\,(H^{\dag}H), \enskip
(e_c d_c)\,u_c^2\,(H^{\dag}H).
\end{gathered}
\end{equation}

\noindent \underline {Class  $\mathcal O(\psi^4\, X)$}: There are 78 operators with four fermions and a single field strength (all plus $\hc$):
\begin{equation}
\begin{gathered}
(d^{\dag}_c\, d_c)(e^{\dag}_c\, e_c) B^L, \enskip
(u^{\dag}_c\, u_c)(e^{\dag}_c\, e_c) B^L, \enskip
2\, (d^{\dag}_c\, d_c)(u^{\dag}_c\, u_c) B^L, \enskip
(d^{\dag}_c\, d_c)(L^{\dag}\, L) B^L, \enskip
(u^{\dag}_c\, u_c)(L^{\dag}\, L) B^L, \\
(e^{\dag}_c\, e_c)(L^{\dag}\, L) B^L, \enskip
(e^{\dag}_c\, e_c)(Q^{\dag}\, Q) B^L, \enskip
(d_c Q)(e^{\dag}_c\, L^{\dag})B^L, \enskip
(d_c Q)(e^{\dag}_c\, L^{\dag})B^R, \enskip
2\, (L^{\dag}\, L)(Q^{\dag}\, Q) B^L, \\
2\,(d^{\dag}_c\, d_c)(Q^{\dag}\, Q) B^L, \enskip
2\,(u^{\dag}_c\, u_c)(Q^{\dag}\, Q) B^L, \enskip
3\, (e_c\, L)(u_c\, Q)B^L, \enskip
3\, (u_c\, d_c)\, Q^2\, B^L, \enskip
(d^{\dag}_c\, d_c)(L^{\dag}\, L) W^L, \\
(e^{\dag}_c\, e_c)(L^{\dag}\, L) W^L, \enskip
(e^{\dag}_c\, e_c)(Q^{\dag}\, Q) W^L, \enskip
(u^{\dag}_c\, u_c)(L^{\dag}\, L) W^L, \enskip
(L^{\dag}\,L)^2 W^L, \enskip
(e^{\dag}_c\, L^{\dag})(d_c\, Q)W^L, \\
(e_c\, L)(d^{\dag}_c\, Q^{\dag})W^L, \enskip
2\,(d^{\dag}_c\, d_c)(Q^{\dag}\, Q) W^L, \enskip
2\,(u^{\dag}_c\, u_c)(Q^{\dag}\, Q) W^L, \enskip
3\, (L^{\dag}\, L)(Q^{\dag}\, Q) W^L, \enskip
2\, (Q^{\dag}\,Q)^2 W^L, \\
3\, (e_c\, L)(u_c\, Q)W^L, \enskip
3\, (u_c\, d_c)\,Q^2\,W^L, \enskip
(d^{\dag}_c)^2\,d^2_c\,G^L, \enskip
(u^{\dag}_c)^2\,u^2_c\,G^L, \enskip
(d^{\dag}_c\, d_c)(e^{\dag}_c\, e_c)\,G^L, \\
(u^{\dag}_c\, u_c)(e^{\dag}_c\, e_c)\,G^L, \enskip
4\,(d^{\dag}_c\, d_c)(u^{\dag}_c\, u_c)\,G^L, \enskip
(Q^{\dag}\, Q)(e^{\dag}_c\, e_c)\,G^L, \enskip
(d^{\dag}_c\, d_c)(L^{\dag}\, L)\,G^L, \enskip
(u^{\dag}_c\, u_c)(L^{\dag}\, L)\,G^L, \\
2\, (Q^{\dag}\, Q)(L^{\dag}\, L)\,G^L, \enskip
4\, (d^{\dag}_c\, d_c)(Q^{\dag}\, Q)\,G^L, \enskip
4\, (u^{\dag}_c\, u_c)(Q^{\dag}\, Q)\,G^L, \enskip
2\, (Q^{\dag})^2\,Q^2 G^L, \enskip
(d_c\, Q)(e^{\dag}_c\, L^{\dag})\,G^L, \\
(d_c\, Q)(e^{\dag}_c\, L^{\dag})\,G^R, \enskip
3\, (e_c\, L)(u_c\, Q)\, G^L, \enskip
6\, (d_c\, u_c)\,Q^2\, G^L .
\end{gathered}
\end{equation}
Additionally, this class contains 22 baryon-number-violating operators (all plus $\hc$):
\begin{equation}
\begin{gathered}
L\, Q^3\, B^L, \enskip
(u_c\, d_c)(L^{\dag}\, Q^{\dag})B^L, \enskip
(u_c\, d_c)(L^{\dag}\, Q^{\dag})B^R, \enskip
(e_c\, u_c)(Q^{\dag})^2\, B^L, \enskip
2\,(e_c\, d_c)\,u^2_c\, B^L, \\
2\,L\, Q^3\, W^L, \enskip
(u_c\, d_c)(L^{\dag}\, Q^{\dag})W^L, \enskip
(u_c\, d_c)(L^{\dag}\, Q^{\dag})W^R, \enskip
(e^{\dag}_c\, u^{\dag}_c)\,Q^2\, W^L, \\
2\, L\, Q^3\, G^L, \enskip
2\, (u_c\, d_c)(L^{\dag}\, Q^{\dag})G^L, \enskip
2\, (u_c\, d_c)(L^{\dag}\, Q^{\dag})G^R, \\
(e_c\, u_c)(Q^{\dag})^2\, G^L, \enskip
(e_c\, u_c)(Q^{\dag})^2\, G^R, \enskip
3\, (e_c\, d_c)u^2_c\, G^L.
\end{gathered}
\end{equation}

\noindent \underline{Class $\mathcal O(\psi^2\, H^3\, X)$}: This class contains 11 operators (all
plus $\hc$),
\begin{equation}
\begin{gathered}
(e_c\,L)\, H^{\dag}\, (H^{\dag}H)B^L, \enskip
(d_c\,Q)\, H^{\dag}\, (H^{\dag}H)B^L, \enskip
(u_c\,Q)\, H\, (H^{\dag}H)B^L, \enskip
2\,(e_c\,L)\, H^{\dag}\, (H^{\dag}H)W^L, \\ 
2\,(d_c\,Q)\, H^{\dag}\, (H^{\dag}H)W^L, \enskip
2\,(u_c\,Q)\, H\, (H^{\dag}H)W^L, \enskip
(d_c\,Q)\, H^{\dag}\, (H^{\dag}H)G^L, \enskip
(u_c\,Q)\, H\, (H^{\dag}H)G^L.
\end{gathered}
\end{equation}

\noindent \underline{Class $\mathcal O(\psi^2\, H\, X^2)$}: The 48 operators (all plus $\hc$) in this class are:
\begin{equation}
\begin{gathered}
(e_c\, L)\, H^{\dag}\, (B^L)^2, \enskip
(e_c\, L)\, H^{\dag}\, (B^R)^2, \enskip
(d_c\, Q)\, H^{\dag}\, (B^L)^2, \enskip
(d_c\, Q)\, H^{\dag}\, (B^R)^2, \\
(u_c\, Q)\, H\, (B^L)^2, \enskip
(u_c\, Q)\, H\, (B^R)^2, \enskip
2\, (e_c\, L)\,H^{\dag}\, B^LW^L , \enskip
(e^{\dag}_c\,L^{\dag})\, H\, B^L W^L, \\
2\, (d_c\, Q)\,H^{\dag}\, B^LW^L, \enskip
(d^{\dag}_c\,Q^{\dag})\, H\, B^L W^L, \enskip
2\, (u_c\, Q)\,H\, B^LW^L , \enskip
(u^{\dag}_c\,Q^{\dag})\, H^{\dag}\, B^L W^L, \\
2\,(e_c\, L)\, H^{\dag}\, (W^L)^2, \enskip
(e^{\dag}_c\,L^{\dag})\,H\,(W^L)^2, \enskip
2\,(d_c\, Q)\, H^{\dag}\, (W^L)^2 , \enskip
(d^{\dag}_c\,Q^{\dag})\,H\,(W^L)^2 ,\\
2\,(u_c\, Q)\, H\, (W^L)^2 , \enskip
(u^{\dag}_c\,Q^{\dag})\,H^{\dag}\,(W^L)^2, \enskip
2\,(d_c\, Q)\,H^{\dag}\, B^L G^L, \enskip
(d_c\, Q)\,H^{\dag}\, B^R G^R , \\
2\,(d_c\, Q)\,H^{\dag}\, W^L G^L, \enskip
(d_c\, Q)\,H^{\dag}\, W^R G^R , \enskip
2\,(u_c\, Q)\,H\, B^L G^L, \enskip
(u_c\, Q)\,H\, B^R G^R, \\
2\,(u_c\, Q)\,H\, W^L G^L, \enskip
(u_c\, Q)\,H\, W^R G^R, \enskip
(e_c\, L)\,H^{\dag}\, (G^L)^2 , \enskip
(e_c\, L)\,H^{\dag}\, (G^R)^2, \\
3\, (d_c\, Q)\, H^{\dag}\, (G^L)^2 , \enskip
2\, (d_c\, Q)\,H^{\dag}\, (G^R)^2 , \enskip
3\, (u_c\, Q)\, H\, (G^L)^2, \enskip
2\, (u_c\, Q)\,H\, (G^R)^2. 
\end{gathered}
\end{equation}

\noindent \underline{Class $\mathcal O(H^4\, X^2)$}: there are 5 operators (all
plus $\hc$) in this class:
\begin{equation}
\begin{gathered}
(H^{\dag}H)^2(B^L)^2, \enskip
(H^{\dag}H)^2\,B^L W^L, \enskip
2\,(H^{\dag}H)^2\,(W^L)^2, \enskip
(H^{\dag}H)^2(G^L)^2.
\end{gathered}
\end{equation}

\noindent\underline{Class $\mathcal O(H^2\, X^3)$}: Here we find  3 operators (all plus $\hc$):
\begin{equation}
\begin{gathered}
(H^{\dag}H) B^L (W^L)^2, \enskip
(H^{\dag}H) (W^L)^3, \enskip
(H^{\dag}H) (G^L)^3,
\end{gathered}
\end{equation}

\noindent \underline{Class $\mathcal O(X^4)$}: The final class contains 9 self-Hermitian operators
\begin{equation}
\begin{gathered}
(B^L)^2 (B^R)^2, \enskip
2\, (W^L )^2(W^R)^2, \enskip
3\, (G^L)^2(G^R)^2, \\
B^LB^RW^LW^R, \enskip
B^LB^RG^LG^R, \enskip
G^LG^RW^LW^R, 
\end{gathered}
\end{equation}
and 17 operators with Hermitian conjugates
\begin{equation}
\begin{gathered}
(B^L)^4, \enskip
2\,(W^L)^4, \enskip
3\,(G^L)^4, \\
2\,(B^L)^2(W^L)^2, \enskip
(B^R)^2(W^L)^2, \enskip
2\,(B^L)^2(G^L)^2, \enskip
(B^R)^2(G^L)^2, \\
B^L (G^L)^3, \enskip
B^R (G^L)^2 G^R, \enskip
2\,(W^L)^2(G^L)^2, \enskip
(W^L)^2(G^R)^2.
\end{gathered}
\end{equation}

All in all, for $N_f = 1$ there are 257 dimension-8 operators without
derivatives (not counting Hermitian conjugates). The total comes to 471
operators if Hermitian conjugates are included.  Of these 257 operators, 64 are
without field strength tensors, and the remaining 193 contain at least one
$X_{\mu\nu}$. The baryon-number-violating operators at $\mathcal{O}(D^0)$ make
up 28 of the 257 operators.

\subsection{Operators with derivatives: $\mathcal O(D)$}
\label{sec:dim8derivs}

Moving to operators with derivatives, we now have to see which term can we use
with our simple subtraction algorithm Eq.~(\ref{eq:himp2}). We'll begin by
enumerating the operator classes at each derivative order, then checking class
by class. However, as we learned from dimension-7, sometimes we need more detail
than just which fields (scalar, fermion, field strength) compose a particular
class. 

At $\mathcal O(D)$ there are four classes: $\mathcal O(D\,\psi^2\, H^4)$,
$\mathcal O(D\, \psi^4\,H)$, $\mathcal O(\psi^2\, X^2)$ and $\mathcal O(\psi^2\,
H^2\, X)$
\begin{itemize}
\item $\mathcal O(D\,\psi^2\, H^4)$: This can be Lorentz invariant only if the derivative is applied to one of the Higgs fields and the fermions have opposite chirality, thus there is only one Lorentz contraction per partition regardless of internal quantum numbers.
\item $\mathcal O(D\, \psi^4\, H)$: The fermion chiralities must be either LLLR or RRRL, and the derivative must act either on the Higgs field or one of the fermions of the dominant chirality. In either case, there is only one Lorentz contraction, though internal quantum numbers may forbid or augment certain derivative locations\footnote{An example of with fewer derivative possibilities is $D(e^{\dag}_c\, e^2_c\, L\, H)$, where two fermion fields are identical, and example of a combination with more is $D(L^{\dag}\,L\, Q\, d_c\,H).$}.
\item $\mathcal O(D\, \psi^2\, X^2)$: When the gauge fields are $X^LX^R$, the derivative must act on one of the fermions, and the two fermions must have opposite chirality. Each possibility has a single Lorentz contraction. When the field strengths are $(X^L)^2$ or $(X^R)^2$, the fermions still must have opposite chirality. The derivative can act on one of the gauge fields or on the LH fermion (for $(X^L)^2$) or RH fermion (for $(X^R)^2)$, and both choices admit a single Lorentz contraction.
\item $O(\psi^2\, H^2\, X)$: Operators in this class can only be Lorentz singlets if the fermions have opposite chirality and the derivative acts on one of the Higgses. There is only one contraction.
\end{itemize} 

All $\mathcal O(D)$ operators at dimension-8 pass the criteria for
Eq.~(\ref{eq:himp2}). Applying it using Eq.~(\ref{eq:SMPE}) as the argument, the
results are listed below and grouped by class:. \\

\noindent \underline{Class $\mathcal O(D\,\psi^2\, H^4)$}: includes the 11 self-Hermitian operators
\begin{equation}
\begin{gathered}
D (d^{\dag}_c\, d_c\,(H^{\dag}H)^2), \enskip
D (e^{\dag}_c\, e_c\,(H^{\dag}H)^2), \enskip
D (u^{\dag}_c\, u_c\,(H^{\dag}H)^2), \\
4\,D (L^{\dag}\, L\,(H^{\dag}H)^2), \enskip
4\,D (Q^{\dag}\, Q\,(H^{\dag}H)^2),
\end{gathered}
\end{equation}
and a single operator with a Hermitian conjugate
\begin{equation}
D (d^{\dag}_c\, u_c\, H^3 H^\dag) .
\end{equation}

\noindent \underline{Class $\mathcal O(D\, \psi^4\, H)$}: this class contains 67 operators (all plus $\hc$):
\begin{equation}
\begin{gathered}
3\, D (d^{\dag}_c d_c\,L\, e_c\,H^{\dag}), \enskip
D (e^{\dag}_c e_c\,L\, e_c\,H^{\dag}), \enskip
3\, D (L^{\dag} L^2\, e_c\,H^{\dag}), \enskip
3\, D (d^{\dag}_c\, d^2_c\, Q\, H^{\dag}), \\
3\, D (e^{\dag}_c e_c\, Q\,d_c\,H^{\dag}), \enskip
6\, D (L^{\dag} L\,Q\,d_c\,H^{\dag}), \enskip
6\, D (Q^{\dag} Q\,L\, e_c\,H^{\dag}), \enskip
6\, D (Q^{\dag} Q^2\,d_c\,H^{\dag}), \\
3\, D (d^{\dag}_c u_c\,L\,e_c\, H), \enskip
6\, D (d^{\dag}_c d_c\,Q\,u_c\,H ), \enskip 
3\, D (e^{\dag}_c e_c\,Q\,u_c\,H),\enskip 
6\, D (L^{\dag} L\,Q\,u_c\,H), \\
6\, D (Q^{\dag} Q^2\, u_c\,H), \enskip
3\, D (u^{\dag}_c u_c\,L\,e_c\,H^{\dag}), \enskip
6\, D (u^{\dag}_c u_c\,Q\, d_c\,H^{\dag}), \enskip
3\, D (u^{\dag}_c u^2_c\,Q\,H).
\end{gathered}
\end{equation}
as well as 16 baryon-number-violating operators (all
plus $\hc$):
\begin{equation}
\begin{gathered}
3\, D (d^{\dag}_c\, Q^2\, L\, H), \enskip
D (e^{\dag}_c\ Q^3\,H), \enskip
2\, D (d^2_c\, u_c\, L^{\dag}\, H^{\dag}), \enskip
3\, D (d_c\, e_c\, Q^{\dag}\, u_c\, H^{\dag}), \\
3\, D ((Q^{\dag})^2\, L^{\dag}\, u_c\,H), \enskip
2\, D (L^{\dag}\, u_c^2\, d_c\, H), \enskip
2\, D (Q^{\dag}\, u^2_c\, e_c\, H) .
\end{gathered}
\end{equation}

\noindent \underline{Class $\mathcal O(D\, \psi^2\, X^2)$}: this class contains  23 self-Hermitian operators:
\begin{equation}
\begin{gathered}
D (e^{\dag}_c e_c\, B^L B^R), \enskip
D (d^{\dag}_c d_c\, B^L B^R), \enskip
D (u^{\dag}_c u_c\, B^L B^R), \enskip
D (Q^{\dag} Q\, B^L B^R), \\
D (L^{\dag} L\, B^L B^R), \enskip
3\,D (d^{\dag}_c d_c\, G^L G^R), \enskip
D (e^{\dag}_c e_c\, G^L G^R), \enskip
D (L^{\dag} L\, G^L G^R), \\
3\,D (Q^{\dag} Q\, G^L G^R), \enskip
3\,D (u^{\dag}_c u_c\, G^L G^R), \enskip 
D (d^{\dag}_c d_c\, W^L W^R), \enskip 
D (e^{\dag}_c e_c\, W^L W^R), \\ 
D (u^{\dag}_c u_c\, W^L W^R), \enskip
2\, D (L^{\dag} L\, W^L W^R), \enskip 
2\, D (Q^{\dag} Q\, W^L W^R), 
\end{gathered}
\end{equation}
and 17 operators with Hermitian conjugates
\begin{equation}
\begin{gathered}
D (d^{\dag}_c d_c\, B^L G^L), \enskip
D (d^{\dag}_c d_c\, B^R G^L), \enskip
D (d^{\dag}_c d_c\, (G^L)^2), \enskip 
D (Q^{\dag} Q\, B^L G^L), \\
D (Q^{\dag} Q\, B^R G^L), \enskip
D (Q^{\dag} Q\, (G^L)^2), \enskip 
D (u^{\dag}_c u_c\, B^L G^L), \enskip
D (u^{\dag}_c u_c\, B^R G^L), \\ 
D (u^{\dag}_c u_c\, (G^L)^2), \enskip
D (L^{\dag} L\, B^L W^L), \enskip 
D (L^{\dag} L\, B^R W^L), \enskip
D (Q^{\dag} Q\, B^L W^L, \\
D (Q^{\dag} Q\, B^R W^L), \enskip 
D (Q^{\dag} Q\, G^L W^L), \enskip
D (Q^{\dag} Q\, G^R W^L), \enskip 
D (L^{\dag} L\, (W^L)^2), \\ 
D (Q^{\dag} Q\, (W^L)^2).
\end{gathered}
\end{equation}

\noindent \underline{Class $\mathcal O(D\, \psi^2\, H^2\, X)$}: this last class contains 46 operators (all plus $\hc$):
\begin{equation}
\begin{gathered}
2\, D (d^{\dag}_c d_c\,H^{\dag}H\,B^L), \enskip
2\, D (e^{\dag}_c e_c\,H^{\dag}H\,B^L), \enskip 
2\, D (d^{\dag}_c d_c\,H^{\dag}H\,G^L), \enskip 
4\, D (L^{\dag} L\,H^{\dag}H\,B^L), \\
4\, D (Q^{\dag} Q\,H^{\dag}H\,B^L), \enskip 
4\, D (Q^{\dag} Q\,H^{\dag}H\,G^L), \enskip 
D (d^{\dag}_c u_c\,(H)^2\,B^L), \enskip 
D (d^{\dag}_c u_c\,(H)^2\,B^R), \\
D (d^{\dag}_c u_c\,(H)^2\,G^L), \enskip 
D (d^{\dag}_c u_c\,(H)^2\,G^R), \enskip 
2\, D (u^{\dag}_c u_c\,H^{\dag}H\,B^L), \enskip
2\, D (u^{\dag}_c u_c\,H^{\dag}H\,G^L), \\
2\, D (d^{\dag}_c d_c\,H^{\dag}H\,W^L), \enskip 
2\, D (e^{\dag}_c e_c\,H^{\dag}H\,W^L), \enskip 
6\, D (L^{\dag} L\,H^{\dag}H\,W^L), \enskip 
6\, D (Q^{\dag} Q\,H^{\dag}H\,W^L), \\
2\, D (u^{\dag}_c u_c\,H^{\dag}H\,W^L), \enskip
D (d^{\dag}_c u_c\,(H)^2\,W^L), \enskip 
D (d^{\dag}_c u_c\,(H)^2\,W^R).
\end{gathered}
\end{equation}

Summing up all of the contributions gives a total of 181 dimension-8 operators at
$\mathcal{O}(D)$. If Hermitian conjugates are included, the total is 328
operators.

\subsection{Operators with derivatives: $\mathcal O(D^2)$}

We continue the procedure at $\mathcal O(D^2)$. There are now seven classes: $\mathcal O(D^2\, (H\, H^{\dag})^3)$, $\mathcal O(D^2\, X^3)$, $\mathcal O(D^2\, \psi^4)$, $\mathcal O(D^2\, \psi^2\, H^3)$, $\mathcal O(D^2\, H^2\, X^2)$, $\mathcal O(D^2\, \phi^4\, X)$ and $\mathcal O(D^2\, \psi^2\, H\, X)$:
\begin{itemize}
\item $\mathcal O(D^2\, H^3\, H^{\dag3})$: This class is the same as the complex scalar field we considered earlier. At $O(D^2)$ all partitions have only one Lorentz contraction.
\item $\mathcal O(D^2\, X^3)$: Either all three field strengths are $X^L$ or $X^R$, or two are $X^L\, (X^R)$ with the third $X^R (X^L)$. In the first case, the partition (1,1,0) is the only way to form a Lorentz invariant, and there is only one contraction. Of course, Bose symmetry or internal quantum numbers may forbid such a possibility, but even when allowed there is only one Lorentz contraction. In the mixed case, i.e $(X^L)^2\,X^R$, we can only form a Lorentz invariant if each of the $X^L$ are acted on with a derivative. This partition has only one contraction.
\item $\mathcal O(D^2\, H^2\, X^2)$: More correctly, this class is $\mathcal O(D^2\, H\,H^{\dag}\, X^2)$. In the case of a product of $X^LX^R$ gauge fields, the derivatives must act on the Higgses. The possibilities are $(2;0;0;0)$,$(0;2;0;0)$, or $(1;1;0;0)$, all of which admit a single Lorentz contraction regardless of how we stitch the $SU(2)_w$ indices. For $(X^L)^2$ or $(X^R)^2$, two derivatives on a single Higgs no longer works. The Lorentz invariant possibilities are $(1;1;0,0)$, $(1;0;1,0)$ and $(0;1;1,0)$, all of which have a single contraction. However, if the two $X^{L,R}$ field strengths are different, i.e. $B^L\, W^L$, the number of partitions grows to 5 and $(1;1;0;0)$ (now $DH^{\dag}\, DH\, X^L_i\, X^L_j$ where $i$ and $j$ label different gauge groups) permits two contractions and therefore must be checked manually.
\item $\mathcal O(D^2\, H^4\, X) = \mathcal O(D^2\, H^2\,H^{\dag 2}\, X)$: The derivatives cannot act on $X$, and two derivatives on a single $H$ or $H^{\dag}$ vanishes by the asymmetry of $X$, so the only possibilities are $(1,1;0,0;0)$, $(0,0;1,1;0)$ and $(1,0;1,0;0)$, all of which can only be Lorentz contracted one way. 
\item  $\mathcal O(D^2\, \psi^4)$; The four fermions must either have LL RR chirality or LLLL/RRRR. In the mixed chirality case there are three partitions. For $(1;0;1;0)$ there is only one Lorentz contraction, so no issue. For $(1;1;0;0)$ (in writing the partition this way we are allowing for the same chirality fermions to be different), $D\psi_{L,i}\,D\psi_{L,j} = (0 \oplus 1, 0 \oplus 1 \oplus 2)$ under ($SU(2)_R, SU(2)_L, SU(2)_w$) and $\psi_{Rk}\psi_{Rl} = (0 \oplus 1,0)$, so there are two different contractions. This means Eq.~(\ref{eq:himp2}) is not guaranteed and we have to look at the internal quantum numbers. For instance, if the two RH fermions (in this example) are identical, the $(1,0)$ (of $SU(2)_R \otimes SU(2)_L)$ component of their product is forbidden unless they carry some other asymmetric quantum number. The partition (0,0; 1,1) has the same issue. For LLLL/RRRR chirality operators, all partitions allow multiple contractions.

\item  $\mathcal O(D^2\, \psi^2\, H^3)$: This class is actually $\mathcal O(D^2 \psi_{Li}\psi_{Li}\, H^2\, H^{\dag})$ (+ h.c.), with one of the two $\psi$ transforming as an $SU(2)_w$ doublet (the two $\psi$ are then obviously different fields). There are many possible partitions, but the problematic one is $\psi_{Li}\psi_{Lj}\, DH\,H\, DH^{\dag} = (0;0;1,0;1)$. For this partition, the product of $\psi_{Li}\psi_{Lj}\, H$ sits in the $(0, 0 \oplus 1; 0 \oplus 1)$ representation of $SU(2)_R \otimes SU(2)_R \otimes SU(2)_w$~\footnote{Expanded out, this stands for $(0,0;0) \oplus (0,0;1) \oplus (0,1;0) \oplus (0,1;1)$}, while the rest of the operator $DH\, DH^{\dag}$ is in the $(0 \oplus 1, 0 \oplus 1; 0 \oplus 1)$ representation. Combining the two pieces, we see that there are two Lorentz contractions for each $SU(2)_w$ assignment (i.e. the $SU(2)_w$ singlet part of $\psi_{Li}\psi_{Lj}\,H$ can be either a Lorentz singlet or triplet when combining with $DH\,DH^{\dag}$). Thus, we cannot guarantee Eq.~(\ref{eq:himp2}) will work for this class of operators.
\item  $\mathcal O(D^2\, \psi^2\, H\, X)$: As with the term above, both fermions must have the same chirality, represent different fields, and one must be an $SU(2)_w$ doublet. Also like the term above, this term fails the criteria for Eq.~(\ref{eq:himp2}). The problematic partition is $D\psi_{Li}\, \psi_{Lj}\, DH\, X$; both the derivative and non-derivative portions lie in Lorentz representation $(\frac 1 2, \frac 1 2 \oplus \frac 3 2)$ (assuming $X^L$, the same logic goes through for $X^R$), and we can make Lorentz invariants by contracting either $SU(2)_L$ possibility regardless of any $SU(2)_w$ quantum numbers.
\end{itemize}

Recapping, four of the operator classes, $\mathcal O(D^2\, H^3\, H^{\dag3})$,
$\mathcal O(D^2\, X^3)$, $\mathcal O(D^2\, H^2\, X^2)$, and $\mathcal O(D^2\,
H^4\, X)$ satisfy the criteria for Eq.~(\ref{eq:himp2}) and we can therefore
automatically determine the number of invariants including IBP redundancies.\footnote{With
	the exception of $D^2(H^{\dag}H\,B^{L,R}W^{L,R})$} Of the remaining
	classes, $\mathcal O(D^2\, \psi^2\, H^3)$ and $\mathcal O(D^2\, \psi^2\,
	H\, X)$ fail for any choice of fermion or gauge fields, while operators
	$\mathcal O(D^2\, \psi^4)$ may be okay depending on the internal
	symmetry quantum numbers of the fermions. For the operators classes
	where Eq.~(\ref{eq:himp2}) does not apply, the number of invariants can
	still be constructed, but it must be done manually by constructing the
	matrix of constraints from $\h_{D,SMEFT}$ and determining its rank. For
	the classes where Eq.~(\ref{eq:himp2}) applies, we have the following
	two-derivative dimension-8 operators.\\

\noindent\underline{Class $\mathcal O(D^2\, H^3\, H^{\dag3})$}: this class contains 2 self-Hermitian operators at $\mathcal{O}(D^2H^6)$
\begin{align}
2\,D^2 (H^{\dag}H)^3.
\end{align}

\noindent \underline{Class $\mathcal O(D^2\, X^3)$}: after IBP redundancies have been considered, we find no operators of this type.\\

\noindent \underline{Class $\mathcal O(D^2\, H^2\, X^2)$}: this class contains 4 self-Hermitian operators:
\begin{equation}
D^2 (H^{\dag}H\,B^L B^R), \enskip
D^2 (H^{\dag}H\, G^L G^R), \enskip
2\, D^2 (H^{\dag}H\, W^L W^R),
\end{equation}
and 5 operators with Hermitian conjugates
\begin{equation}
\begin{gathered}
D^2 (H^{\dag}H\,(B^L)^2), \enskip
D^2 (H^{\dag}H\,(G^L)^2),  \enskip
D^2 (H^{\dag}H\,(W^L)^2), \\
D^2 (H^{\dag}H\,B^L W^L), \enskip
D^2 (H^{\dag}H\, B^R W^L). 
\end{gathered}
\end{equation}

\noindent \underline{Class $\mathcal O(D^2\, H^2\,H^{\dag 2}\, X)$}: there is just one operator in this class (plus $\hc$)
\begin{align}
D^2 ((H^{\dag}H)^2 W^L),
\end{align}

Next we can list the operators from the remaining classes. As stated above, the
number of independent invariant operators in these classes must be determined
manually.\footnote{Although listed earlier, $D^2(H^{\dag}H\, B^{L,R}W^{L,R})$ is also in this operator set.} \\

\noindent \underline{Class  $\mathcal O(D^2\, \psi^4)$}: this class contains 39
self-Hermitian operators

\begin{equation}
\begin{gathered}
2\, D^2 (d^{\dag}_c d_c)^2, \enskip
2\, D^2 (u^{\dag}_c u_c)^2, \enskip
4\, D^2 (Q^{\dag} Q)^2,     \enskip
2\, D^2 (L^{\dag} L)^2,     \enskip 
D^2 (e^{\dag}_c e_c)^2,  \\
2\, D^2 (d^{\dag}_c d_c\, e^{\dag}_c e_c), \enskip
2\, D^2 (u^{\dag}_c u_c\, e^{\dag}_c e_c), \enskip
2\, D^2 (L^{\dag} L\, e^{\dag}_c e_c), \enskip
2\, D^2 (Q^{\dag} Q\, e^{\dag}_c e_c), \\
2\, D^2 (L^{\dag} L\, d^{\dag}_c d_c), \enskip
2\, D^2 (L^{\dag} L\, u^{\dag}_c u_c), \enskip
4\, D^2 (Q^{\dag} Q\, d^{\dag}_c d_c), \enskip
4\, D^2 (Q^{\dag} Q\, u^{\dag}_c u_c), \\
4\, D^2 (u^{\dag}_c u_c\, d^{\dag}_c d_c), \enskip
4\, D^2 (Q^{\dag} Q\, L^{\dag} L)
\end{gathered}
\end{equation}
and 6 operators with Hermetian conjugate. These can be grouped into 4 which
preserve baryon number
\begin{equation}
2\, D^2 (L^{\dag} e^{\dag}_c\, Q\, d_c), \enskip
2\, D^2 (e_c\, L\, u_c\, Q), \\
\end{equation}
and 2 that do not.
\begin{equation}
2\, D^2 (L^{\dag}\, Q^{\dag}\, u_c\, d_c), \enskip
\end{equation}
Operators in this class containing fermions of mixed chirality (LLRR) and
multiple powers of the same field (i.e. $d^2_c$) actually satisfy the criteria
for Eq.~(\ref{eq:himp2}) so the number of invariants can be determined
automatically.  This occurs because Fermi statistics limits the representations,
thereby removing the troublesome extra Lorentz contractions. \\

\noindent \underline{Class $\mathcal O(D^2\, \psi^2\, H^3)$}: we find 15 operators in this class (all plus $\hc$):
\begin{equation}
5\, D^2 (L\, e_c\, H\,(H^{\dag})^2), \enskip
5\, D^2 (Q\, d_c\, H\,(H^{\dag})^2), \enskip
5\, D^2 (Q\, u_c\, H^2\,H^{\dag}).
\end{equation}

\noindent \underline{Class $\mathcal O(D^2\, \psi^2\, H\, X)$}: If the fermion's chirality is different than the gauge field, i.e $\psi_L\psi_L X^R$, there is only one contraction per partition and Eq.~(\ref{eq:himp2}) holds. For the `same chirality' combinations, the invariants must be determined manually. We find 8 terms in this category, all +h.c.
\begin{equation}
\begin{gathered}
D^2\, (Q\, d_c\, H^{\dag}\, F^L), \enskip
D^2\, (Q\, d_c\, H^{\dag}\, G^L), \enskip
D^2\, (Q\, d_c\, H^{\dag}\, W^L), \enskip
D^2\, (Q\, u_c\, H\, F^L), \\
D^2\, (Q\, u_c\, H\, G^L), \enskip
D^2\, (Q\, u_c\, H\, W^L), \enskip
D^2\, (L\, e_c\, H^{\dag}\, F^L), \enskip
D^2\, (L\, e_c\, H^{\dag}\, W^L), \enskip
\end{gathered}
\end{equation}

\subsection{Operators with derivatives: $\mathcal O(D^3)$}

As we increase the number of derivatives, the number of operator classes shrinks. At $\mathcal O(D^3)$ there are only two possibilities, $\mathcal O(D^3\psi^3X)$ and $\mathcal O(\psi^3 H^2)$.
\begin{itemize}
\item $\mathcal O(D^3\,\psi^2\,X)$: To be Lorentz invariant the fermions must have opposite chirality. There are three ways to partition the derivatives, depending on whether the gauge field is $X^L$ or $X^R$. For $X^L$, one can have two derivatives on $\psi_L$ and one on $\psi_R$,  two derivatives on $\psi_L$ and one on $X^L$, or one derivative on each. For each partition there is a single Lorentz contraction. As always, fermions with internal quantum numbers may have more invariants per contraction.
\item $\mathcal O(D^3\,\psi^2 H^2)$: More correctly, $\mathcal O(D^3\,\psi^2 H\, H^{\dag})$, with the fermions having opposite chirality. There are 10 different partitions, most of which only allow one contraction, however the partitions $D\psi_L\, \psi_R\, DH\, DH^{\dag}$ and $\psi_L\, D\psi_R\, DH\, DH^{\dag}$ admit multiple. This class of operator therefore fails the criteria for Eq.~(\ref{eq:himp2}) and the number invariants must be calculated by hand.
\end{itemize}

Following the pattern of the $\mathcal O(D^2)$ operators, we list the operators in the classes satisfying Eq.~(\ref{eq:himp2}) first, then show the operators in classes that must be calculated manually.\\

\noindent \underline{Class $\mathcal O(D^3\,\psi^2\,X)$}: after accounting for all redundancies, we find no operators of this type. \\

\noindent \underline{Class $\mathcal O(D^3\,\psi^2\,H^2)$}: we find 14 self-hermitian operators of this type:
\begin{equation}
\begin{gathered}
2\, D^3\, (d_c\, d^{\dag}_c\, HH^{\dag}),\enskip
2\, D^3\, (u_c\, u^{\dag}_c\, HH^{\dag}),\enskip
2\, D^3\, (e_c\, e^{\dag}_c\, HH^{\dag}),\\
4\, D^3\, (L\, L^{\dag}\, HH^{\dag}),\enskip
4\, D^3\, (Q\, Q^{\dag}\, HH^{\dag})\enskip
\end{gathered}
\end{equation}

\subsection{Operators with derivatives: $\mathcal O(D^4)$}

In this last category there are four classes: $\mathcal O(D^4 X^2)$, $\mathcal
O(D^4 H^{\dag}H\, X)$, $\mathcal O(D^4 \psi^2 H)$ and $\mathcal O(D^4 H^2\, H^{\dag2})$:
\begin{itemize}
\item $\mathcal O(D^4 X^2)$:  Operators in this class are quadratic in fields, so there is only one way to partition
the derivatives and only one contraction
\item $\mathcal O(D^4 H^{\dag}H\, X)$: To be invariant, the field strength must be hypercharge. In this case, there are three possible partitions:
$(2;2;0)$, $(1;2;1)$, and $(2;1;1)$, each of which admits only one contraction.
\item $\mathcal O(D^4 \psi^2 H)$: To maintain $SU(2)_w \otimes U(1)_Y$ invariance, these operators take the form of the four derivatives acting on the familiar $d=4$ Yukawa terms (i.e. $D^4(L\, e_c\, H)$). There are six partitions: $(2;2;0)$, $(1;1;2)$, $(2;1;1)$, $(1;2;1)$, $(2;0;2)$ and $(0;2;2)$, but only one contraction per partition.
\item $\mathcal O(D^4 H^2\, H^{\dag2})$: These operators are analogous to the $\mathcal O(D^4)$ complex scalar operators we
studied in Sec.~\ref{sec:phistar}. As such, there are multiple contractions per derivative partition.
\end{itemize}

For the first three classes, we can rely on Eq.~(\ref{eq:himp2}) and find {\em no} operators after IBP
redundancies are taken into account. For the last class, $\mathcal O(D^4 H^2\, H^{\dag2})$, the number of invariants must be derived manually.  We
find: \\


\noindent \underline{Class $\mathcal O(D^4 H^2\, H^{\dag2})$}: 

\begin{equation}
3\, D^4 (H^2\, H^{\dag2}).
\end{equation}

\subsection{Total number of dimension-8 operators}

The complete set of operators of dimension 8 is thus the sum of the sets above.
In total, the number of operators we predict at dimension 8 for $N_f = 1$ is
$257 + 181 + 80 + 14 + 3 = 535$ operators. If Hermitian conjugates are included in the
counting, the total comes to $471 + 328 + 115 + 14 + 3 = 931$ operators. Of the 535
operators, 46 violate baryon number. The fact that we need to determine the
number of operators in some classes by hand makes it difficult to repeat the
calculation for $N_f > 1$.

\section{Discussion}
\label{discussion}

In this paper we proposed a method to incorporate derivatives into the Hilbert
series, taking into account EOM and IBP  redundancies. The method consists of
two main steps. First, each spurion in the plethystic exponential is first
dressed with derivatives, $\phi \to \phi + D\phi + D^2\phi\, \cdots$ retaining
only the symmetric pieces of $D^m$ when $m\ge 2$. The dressed PE is then used to
form two separate series upon integration over the Haar measure: i.) the series
of total (Lorentz and gauge) invariants, and ii.) the series of PE terms in the
$(\frac 1 2, \frac 1 2)$ representation of the Lorentz group. The difference of
these two series is the derivative-improved Hilbert series. This treatment of
derivatives and their redundancies was inspired in part by recent work on
Hilbert series in $(0+1)$ spacetime dimension~\cite{berkeley}.  We show that
this method of accounting for EOM and IBP overcounting is not foolproof -- it
fails whenever there are multiple possible Lorentz contractions for a given
combination of derivatives, fields, and internal quantum number assignments.
However, even with this limitation, the
derivative-improved Hilbert series does successfully reproduce the previous
counts of the dimension-6 and -7 SMEFT operators, and is applicable for
determining the majority of the dimension-8 operators. We therefore calculate
the $d = 8, N_f = 1$ list of operators by using a brute-force approach for the
few cases where the derivative-improved series does not work. We also provide
the list of $d=7$, arbitrary $N_f$ SMEFT operators in Appendix \ref{dim_7_3}.

Since the Hilbert series technique and the associated mathematics have only
recently been applied to the problem of calculating operators for effective
field theories, there are a number of potential directions for future research.
First, while our main focus in this work has been the SMEFT, the
`derivative-improved' Hilbert series technique we displayed will work for any
EFT where the relevant degrees of freedom transform linearly under all
symmetries of the theory. Similarly, though we have worked exclusively in four
spacetime dimensions, the technique should apply to EFT in other spacetime
dimension as well, after appropriately changing the (Euclideanized) Lorentz
symmetry.  Another possible application is to include non-dynamical spurions
(coupling matrices or vacuum expectation values) in addition to dynamical
spurions (fields and their derivatives) in the Hilbert series. For example, we
could enlarge the symmetry of the SMEFT to include quark and lepton flavor; the
quark and leptons in the PEF would have to be modified to take into account
their flavor transformations, but we could also add flavor bi-fundamental Yukawa
matrices  to the list of spurions in the PE.  To account for the fact that
Yukawa matrices are couplings and not fields, the Yukawa spurions would not be
dressed with derivatives.

Finally, a complete implementation of derivatives accounting for IBP into the
Hilbert series technique is a major goal of this line of research. This would
remove the constraint of requiring a single Lorentz contraction per derivative
partion (and quantum number assignment), and thus would allow the fully
automated calculation of EFT operators to higher orders in derivatives. Such an
implementation would also likely lead to a deeper understanding of why
Eq.~(\ref{eq:himp2}) works in its realm of applicability.

\acknowledgments
We would like to thank the authors of Ref~\cite{berkeley}, Brian Henning,
Xiaochuan Lu, Tom Melia and Hitoshi Murayama for kindly sharing their results
with us over the last few months.  We would also like to acknowledge Sekhar
Chivukula, Brian Henning, Graham Kribs, Tom Melia and Veronica Sanz for useful
discussions. The work of AM was partially supported by the National Science
Foundation under Grant No.~PHY-1417118.  This research was supported in part by
the Notre Dame Center for Research Computing through computing resources.

\appendix
\section{SMEFT fields and quantum numbers}
\label{sec:smeftqn}

In this appendix we list the set of SMEFT spurions and their representations
under $SU(2)_R$,\, $SU(2)_L$,\, $SU(3)_c$,\, $SU(2)_w$,\,$U(1)_Y$.

\begin{table}[h!]
\begin{center}
\begin{tabular}{c|cc|ccc}  \hline
& $SU(2)_R$ & $SU(2)_L$ & $SU(3)_c$ & $SU(2)_w$& $U(1)_Y$\\ \hline
$H$ & 0 & 0 & 0 & 2 & $\frac 1 2$ \\
$Q$ & 0 & $\frac 1 2$ & 3 & 2 & $\frac 1 6$ \\
$u_c$ &0 &$\frac 1 2$ &  $\bar 3$ & 0 & $-\frac 2 3$ \\
$d_c$ &0 &$\frac 1 2$ &  $\bar 3$ & 0 & $\frac 1 3$  \\
$L$ &0 &$\frac 1 2$ &  0 & 2 & $-\frac 1 2$ \\
$e_c$ &0 &$\frac 1 2$ &  0 & 0 & $+1$ \\  
$B^L$ &0 & 1 & 0  & 0 & 0 \\
$W^L$ &0 & 1 & 0  & 3 & 0 \\
$G^L$ &0 & 1 & 8  & 0 & 0 \\
\end{tabular}
\caption{Fields and their quantum numbers of the SMEFT. The quantum numbers for
conjugate fields can be derived by swapping $SU(2)_R$ and $SU(2)_L$
representations, reversing the sign of the $U(1)_Y$ charge, and replacing
$SU(3)_c$ and $SU(2)_w$ representations with their conjugates.} 
\end{center}
\end{table}

\section{From Hilbert series to operators: adding indices}
\label{sec:indices}

Asking for the invariants formed from two left-handed lepton doublets $L$ (of
the same flavor), two $L^{\dag}$ and two Higgses at dimension 8, the output of
the Hilbert series is:
\begin{equation}
2\, (L^{\dag}\,L)^2 (H^{\dag}H),
\end{equation}
where the coefficient 2 indicates that there are two possible invariants. Since
they are anticommuting fermionic fields, the two $L$ fields must form a totally
antisymmetric combination of $(0,1/2; 0, 1/2, -1/2)$ fields, where the numbers
in parenthesis indicate the representation/quantum numbers under $(SU(2)_R$,
$SU(2)_L$; $SU(3)_c$, $SU(2)_w$, $U(1)_Y)$ respectively:
\begin{equation}
(0,1/2; 0, 1/2, -1/2)^2_{anti} = (0, 0\oplus1; 0, 0\oplus1, -1)_{anti} = (0,0;0,1,-1) \oplus (0,1;0,0,-1),
\end{equation}
where antisymmetry kills the other two possibilities. The $(L^{\dag})^2$ piece
is identical, except for $SU(2)_L \to SU(2)_R$ and hypercharge $+1$ instead of
$-1$.  Combining the lepton pieces with the Higgs piece, we get
\begin{align}
L^2 \otimes L^{\dag2} \otimes (H^{\dag}H) & = [(0,0;0,1,-1) \oplus (0,1;0,0,-1)] \otimes [(0,0;0,1,+1) \oplus (1,0;0,0,+1)]\nonumber \\
& \quad\quad\quad\quad\quad \otimes (0,0; 0, 0\oplus1,0).
\end{align}
The two ways to make total invariants are now straightforward. The Lorentz
triplet pieces from $L^2$ and $(L^{\dag})^2$ are in different $SU(2)$ groups, so
they cannot be joined to form an Lorentz invariant. The Lorentz singlet
combination of $L^2$ and $(L^\dag)^2$ can be either a $SU(2)_w$ singlet,
triplet, or quintuplet, and either the singlet or triplet can form an $SU(2)_w$
invariant with the Higgs pair. Writing out the index contractions explicitly, we
have the two independent operators
\begin{equation}
\begin{gathered}
\LC_{AB}
(\epsilon^{\alpha\beta} L_{\alpha,i} \tau^{Ai}_j   L_{\beta}^j)
(\epsilon^{\dot{\alpha}\dot{\beta}} L^{\dag}_{\dot{\alpha},k} \tau^{Bk}_m
L^{\dag m}_{\dot{\beta}})
(\epsilon^{xy}H^{\dag}_x H_y)  \\
\epsilon_{ABC}\,(\epsilon^{\alpha\beta}L_{\alpha,i}\,\tau^{Ai}_j\,
L_{\beta}^j)\,(\epsilon^{\dot{\alpha}\dot{\beta}}
\epsilon^{km}L^{\dag}_{\dot{\alpha},k} \tau^{Bk}_mL^{\dag m}_{\dot{\beta}})\, 
(H^{\dag}_x\,\tau^{Cx}_y H^y).
\end{gathered}
\label{eq:indices}
\end{equation}
Here the undotted (dotted) Greek indices refer to $SU(2)_L$  ($SU(2)_R$)
indices, which are contracted via the epsilon tensor. We work in a convention
where fundamental $SU(2)$ indices are naturally lowered, and where the $(1,2)$
entry of $\epsilon^{\alpha\beta}$ is +1 (for any $\alpha, \beta$, both dotted
and un-dotted). The $\tau^A$ matrices, with one lower and one upper index, are
the canonical Pauli matrices. The Pauli matrices could be omitted in favor of symmetrized $SU(2)_w$ indices.

\section{Hilbert series for a single scalar field}
\label{sec:singlephi}

Here we give the expanded expression for the derivative-improved Hilbert series
for a single scalar field, $\h_{\phi}$. We will use the complex variable $x$ to
parameterize $SU(2)_L$ and $y$ to parameterize $SU(2)_R$. The Haar measure for
the two $SU(2)$ integrations is
\begin{equation}
\int d\mu_{SU(2)_L}d\mu_{SU(2)_R} = \frac{1}{(2\pi\, i)^2}\oint_{|x|=1} dx \,
\frac{(x^2-1)}{x}\oint_{|y|=1}dy\,\frac{(y^2-1)}{y}.
\end{equation}
Following the discussion in Sec.~\ref{sec:EOM}, the spurion for a single derivative on $\phi$ sits in the $(\frac 1 2, \frac 1
2)$ representation, the symmetric combination of two derivatives acting on
$\phi$ sits in the (1,1) representation, and the pattern continues for higher
derivatives. For $SU(2)$ parameterized by $x$, the characters for the smallest
few representations are:
\begin{table}[H]
\centering
\begin{tabular}{c|c}
1 & singlet \\
$\Big(x + \frac 1 x \Big)$ & doublet \\
$\Big(1 + x^2 + \frac 1 {x^2} \Big)$ & triplet \\
$\Big(x + x^3 + \frac 1 x + \frac 1 {x^3} \Big)$ & quadruplet \\
\end{tabular}
\end{table}
Notice the characters are palindromic in $x \to \frac 1 x$ and the dimension of the
representation is obtained by setting $x \to 1$.

Taking the characters from the table above, we can form the PE for the spurion
$\phi$  dressed with derivatives:
\begin{align}
\text{PE}_{\phi} = &\, \text{PE}\Big[\phi\, \Big( 1 + D\,\Big(x + \frac 1 x \Big)\Big(y +
\frac 1 y \Big) + D^2\,\Big(1 + x^2 + \frac 1 {x^2} \Big)\Big(1 + y^2 +
\frac 1 {y^2} \Big) + \nonumber \\
& \quad\quad D^3\,\Big(x + x^3 + \frac 1 x + \frac 1 {x^3} \Big)\Big(y + y^3 +
\frac 1 y +  \frac 1 {y^3} \Big) + \mathcal O(D^4) + \cdots \Big) \Big].
\label{expansion}
\end{align}
To form the PE, each term in the expanded term bracketed above, $\phi, \phi\,
D\,y\, x, \phi\,D\,\frac y x$, etc. must be separately plugged into
Eq.~(\ref{eq:PE}). It is possible the sum the powers of $D$ into a compact expression, similar to those shown in Ref.~\cite{berkeley} for 0+1 dimensions. For scalars, the derivative sum is:
\begin{equation}
\frac{(1-D^2)\,x^2\, y^2\,\phi}{(D\, x - y)(D\, y - x)(D - x\,y)(D\,x\,y - 1)},
\label{eq:dsum}
\end{equation}
and similar expressions can be obtained for fermions or field strengths. Plugging this expression into Eq.~(\ref{expansion}), we can obtain the PE for scalars to all orders in $D$  -- including EOM redundancies. However, as the expressions are quite lengthy and still contain IBP redundancies, we do not repeat them here\footnote{In 1+1 dimensions, where the Lorentz group is just $U(1)$, the all-orders PE is more compact $\frac{1}{1-\phi}\prod_{n=1}^{\infty}\frac{1}{(1 - \phi\, x^{-n})(1 - \phi\, x^n)}$, where $x$ is the complex number parameterizing $U(1).$}. Finally, we find there is an alternative representation of scalar operators in terms of graphs. Specifically, each $\phi$ is a vertex and each contracted pair of derivatives is an edge. As such, the the number of operators with $n$ fields and $m$ derivatives (minus EOM) is equal to the number of loopless multigraphs with $n$ vertices and $m/2$ edges~\cite{Gross:2013:HGT:2613412}.  

%

\section{Dimension-7 operators for arbitrary $N_f$}
\label{dim_7_3}

This appendix lists our prediction for dimension-7 operators produced by the
Hilbert series technique of Eq.~(\ref{eq:himp2}) for arbitrary $N_f$. To our
knowledge, this has not been shown previously. First, we list the
derivative-free operators (all plus $\hc$):
\begin{equation}
\begin{gathered}
\frac{1}{2} \left(N_f^4 - N_f^3 \right) Q (d_c^\dag)^2 e_c H^\dag, \enskip
\frac{1}{3} \left(N_f^4 - N_f^2 \right) L d_c^3 H^\dag, \enskip
N_f^4\, L (Q^\dag )^2 d_c H, \\
N_f^4\, L d_c^2 u_c H, \enskip
N_f^4\, L e_c^\dag u_c^\dag d_c H, \enskip
\frac{1}{2} \left(N_f^2 + N_f\right)  L^2 (H^\dag H) H^2, \\
\frac{1}{2} \left(N_f^2 - N_f\right)  L^2 H^2 B^L, \enskip 
N_f^2\,  L^2 H^2 W^L, \enskip
N_f^4\, L^2 Q^\dag u_c^\dag H, \\
2 N_f^4\, L^2 Q d_c H, \enskip
\frac{1}{2} \left(2 N_f^4 + N_f^2\right) L^3 e_c H.
\end{gathered}
\end{equation}
Next, there are the one and two derivative operators (all plus $\hc$):
\begin{equation}
\begin{gathered}
\frac{1}{6} \left( N_f^4 + 3 N_f^3 + 2 N_f^2 \right) D e_c^\dag d_c^3, \enskip
N_f^2\,  D L e_c^\dag H^3, \enskip
\frac{1}{2} \left(N_f^4 + N_f^3\right) D L Q^\dag d_c^2, \\
\frac{1}{2} \left(N_f^4 + N_f^3\right) D L^2 d_c u_c^\dag, \enskip
\frac{1}{2} \left(N_f^2 + 3 N_f \right) D^2 L^2 H^2.
\end{gathered}
\end{equation}
Adding up all of the contributions, the total number of dimension-7 operators
for arbitrary $N_f$ is (not counting Hermitian conjugates):
\begin{equation}
\frac{1}{6} \left( 52 N_f^4 + 6 N_f^3 + 23 N_f^2 + 9 N_f \right),
	\label{}
\end{equation}
which is 15 for $N_f = 1$ and 768 for $N_f = 3$.

\bibliography{Notes}
\bibliographystyle{jhep}

\end{document}